\documentclass[manuscript,screen,review,acmsmall]{acmart}
\usepackage[noend]{algorithm2e}
\usepackage{algpseudocode}% http://ctan.org/pkg/algorithmicx
\usepackage{graphicx,balance,multirow,xspace,subcaption,longtable}
\usepackage{tabularx,multirow,booktabs,blindtext}
\usepackage{makecell,amsmath,cleveref,rotating}
\usepackage[font={small}, textfont=md]{caption}
\usepackage[T1]{fontenc}
\usepackage[utf8]{inputenc}
\usepackage[draft,inline,nomargin,index]{fixme}
\usepackage{marginnote}
\usepackage{booktabs}
\usepackage{xcolor}
\usepackage[bb=boondox,bbscaled=.95,cal=boondoxo]{mathalfa} 
\usepackage{adjustbox}
\usepackage{caption}
\usepackage{subcaption}
\usepackage{colortbl}

\setcopyright{none}

\AtBeginDocument{%
  }

\PassOptionsToPackage{hyphens}{url}
\usepackage{hyperref} 
\newcolumntype{P}[1]{>{\arraybackslash}p{#1}}
\newcolumntype{X}[1]{>{\centering\arraybackslash}p{#1}}

\expandafter\def\expandafter\UrlBreaks\expandafter{\UrlBreaks%  save the current one
  \do\a\do\b\do\c\do\d\do\e\do\f\do\g\do\h\do\i\do\j%
  \do\k\do\l\do\m\do\n\do\o\do\p\do\q\do\r\do\s\do\t%
  \do\u\do\v\do\w\do\x\do\y\do\z\do\A\do\B\do\C\do\D%
  \do\E\do\F\do\G\do\H\do\I\do\J\do\K\do\L\do\M\do\N%
  \do\O\do\P\do\Q\do\R\do\S\do\T\do\U\do\V\do\W\do\X%
  \do\Y\do\Z}

\crefformat{section}{\S#2#1#3}
\crefformat{subsection}{\S#2#1#3}
\crefformat{subsubsection}{\S#2#1#3}

\hypersetup{%
  pdftitle={Uncovering the Interaction Equation},
  pdfauthor={},
  pdfkeywords={},
  bookmarksnumbered,
  pdfstartview={FitH},
  colorlinks,
  citecolor=black,
  filecolor=black,
  linkcolor=black,
  urlcolor=linkColor,
  breaklinks=true,
  hypertexnames=false
}

\newcommand\clearrow{\global\let\rowmac\relax}

\newcommand{\eg}{e.g.,\ }
\newcommand{\etal}{et al.\xspace}
\newcommand{\ie}{i.e.,\ }

\newcommand{\cf}{\emph{cf.\ }}

\newcommand{\para}[1]{\vspace{.05in}\noindent\textbf{#1}}
\newcommand{\paraem}[1]{\vspace{.05in}\noindent\emph{#1}}
\newcommand{\yt}{YouTube\xspace}
\newcommand{\X}{$\mathbb{X}$\xspace}

\copyrightyear{2019} 
\acmYear{2018} 
\setcopyright{acmlicensed}
\acmConference[IMC '19]{acm}{imc}

\setcopyright{none}

\begin{document}

\renewcommand\footnotetextcopyrightpermission[1]{} % removes footnote with conference information in first column
\pagestyle{plain} % removes running headers

\renewcommand{\sectionautorefname}{\S}
\renewcommand{\subsectionautorefname}{\S}
\renewcommand{\subsubsectionautorefname}{\S}

\title{Uncovering the Interaction Equation}
\subtitle{Quantifying the Effect of User Interactions on Social Media Homepage Recommendations}

\author{Hussam Habib}
\affiliation{%
  \institution{University of Iowa}
  \city{Iowa City}
  \country{USA}}
\email{hussam-habib@uiowa.edu}

\author{Ryan Stoldt}
\affiliation{%
  \institution{Drake University}
  \city{Des Moines}
  \country{USA}}
\email{ryan.stoldt@drake.edu}

\author{Raven Maragh-Lloyd}
\affiliation{%
  \institution{Washington University in St. Louis}
  \city{St. Louis}
  \country{USA}}
\email{lloydr@wustl.edu}

\author{Brian Ekdale}
\affiliation{%
  \institution{University of Iowa}
  \city{Iowa City}
  \country{USA}}
\email{brian-ekdale@uiowa.edu}

\author{Rishab Nithyanand}
\affiliation{%
  \institution{University of Iowa}
  \city{Iowa City}
  \country{USA}}
\email{rishab-nithyanand@uiowa.edu}

\renewcommand{\shortauthors}{Habib et al.}
\begin{abstract}
Social media platforms depend on algorithms to select, curate, and deliver content personalized for their users. 
These algorithms leverage users' past interactions and extensive content libraries to retrieve and rank content that personalizes experiences and boosts engagement. 
Among various modalities through which this algorithmically curated content may be delivered, the homepage feed is the most prominent. 
This paper presents a comprehensive study of how prior user interactions influence the content presented on users' homepage feeds across three major platforms: \yt, Reddit, and \X (formerly Twitter).
We use a series of carefully designed experiments to gather data capable of uncovering the influence of specific user interactions on homepage content. 
This study provides insights into the behaviors of the content curation algorithms used by each platform, how they respond to user interactions, and also uncovers evidence of deprioritization of specific topics.
\end{abstract}


\maketitle
\sloppy
\section{Introduction}

\para{Social media platforms have fundamentally transformed the information ecosystem.} 
Recent surveys have shown that roughly eight-in-ten US adults consume content \cite{Shearer2024HowInstagram} on social media platforms and five-in-ten use it for news consumption \cite{Wang2024WhatCenter}. 
This shift marks a significant change from a decade ago when traditional media was the dominant and preferred medium for media and news consumption \cite{Nielsen2014TheNews}.
Unlike traditional media, which delivers information passively, social media allows for interactive and personalized engagement. Users no longer just receive information; they actively like, share, and subscribe to content. 
This participation allows users to provide feedback that shapes the production of information \cite{Bruns2008BlogsProdusage} while also empowering them to alter the reach, or {virality}, of information \cite{Wang2020HowEra}.

\para{The algorithms driving the changing information ecosystem are not fully understood.}
At the root of the increased engagement on social media platforms, are sophisticated algorithms which retrieve and rank selections of the platform's content \cite{Ge2014Patent, Joseph2012Patent}. 
Despite their effectiveness in boosting user engagement, the inner workings of content curation algorithms remain largely unknown to their users. This has necessitated folk theories \cite{DeVito2017AlgorithmsMedia} and research \cite{Mousavi2024AuditingExplanations} to understand their operation.
Further, this opacity has also led to concerns about the potential for algorithmic bias and manipulation. In fact, prior research and whistle-blower documents \cite{FacebookPapers} have shown that content curation algorithms may have (exploitable) tendencies to promote misinformation \cite{Shao2016Hoaxy:Misinformation, habib_morbid_2023}, propaganda \cite{Golovchenko2020Cross-PlatformElection}, ideologically congruent \cite{habib_algorithmic_2024}, and distressing content \cite{Tiggemann2014NetTweens:Girls} while also suppressing content from minority communities \cite{Matamoros-Fernandez2017PlatformedYouTube}. 

\para{User interactions are a key input to content curation algorithms.}
Affordances \cite{Gibson1983TheSystems}, such as likes, shares, and searches, provide mechanisms for users to shape their self-presentation and engage with content, thereby signaling their content consumption habits and preferences \cite{Theocharis2023PlatformEngagement, Bucher2017ThePlatforms}.
These interactions enable content curation algorithms to rank and retrieve content that spurs further user engagement on the platform.
One effective technique, {\em collaborative filtering}, uses these interaction signals to create a network of users with overlapping preferences, from which content recommendations are generated  \cite{Goldberg1992UsingTapestry}.
Consequently, examining how platforms interpret and respond to each affordance-enabled preference signal is crucial for improving our understanding of content curation algorithms and uncovering their potentially problematic tendencies.
Despite their importance, there is little platform audit research that sheds light into how these preference signals shape the content curation process. 
\emph{This paper bridges this gap in knowledge by systematically quantifying the impact of user interactions on the content recommended in platforms' homepage feeds.} 
We use a series of carefully crafted experiments across three major platforms (\Cref{sec:methods}), \yt, Reddit, and \X (formerly Twitter), which share similar affordances yet have different goals, and uncover the specific influence of user interactions on homepage feeds and platform behaviors.
Specifically, we answer the following research questions.
\begin{itemize}
    \item \para{RQ1.} {\em How do user interactions influence the topics shown homepage feeds? (\Cref{sec:rq1})}
    
    We analyze the data gathered from our experiments to understand how each platform's homepage feed algorithm responds to different user engagement signals. 
    Our analysis shows that the homepage feed algorithms of \yt, \X, and Reddit, each interpret user interactions in ways that reflect their platform structures and priorities.
    We see that the topics in a \yt homepage feed are heavily personalized for their users and most strongly influenced by their positive feedback affordance (the Like button) and passive video consumption.
    Reddit's homepage feed also shows heavy personalization which is almost entirely driven by users' interests as indicated by community subscriptions.
    In contrast, \X shows very low to moderate personalization in response to all interactions with the most noticeable, but still moderate, effects occurring due to the Follow interaction.
        
    \item \para{RQ2.} {\em How do user interactions influence platform behaviors? (\Cref{sec:rq2})}

    We examine three different behaviors of each platform's homepage feed curation algorithms: how they navigate the explore-exploit trade-off, their reliance on explicit association interactions (\eg Join and Follow) when selecting and ordering content, and their response to consecutive near-identical interactions from a user.
    We find that \yt's algorithm demonstrates more exploitative tendencies than \X or Reddit and that the algorithms are  primarily driven by specific engagement actions (rather than the topics associated with the engagement).
    We also find that all platforms heavily rely on explicit association interactions from a user when ranking content for their homepage feed and the effect is less dramatic in their content selection algorithms.
    Finally, we find strong evidence of deprioritization of politics-related content on \X. We discover that user interactions with political content or their creators are treated significantly differently by their homepage feed algorithms, than interactions with non-political content.
\end{itemize}
We then discuss the limitations and implications of our study (\Cref{sec:discussion}). Finally, we place our work in the context of prior research (\Cref{sec:related}). 
Taken together, our investigation yields several useful insights into the workings of opaque platforms' homepage feed curation algorithms and how they respond to user interactions.

\section{Background} \label{sec:background}
In this section, we provide an overview of platform affordances (\Cref{sec:background:affordances}), content curation algorithms (\Cref{sec:background:algorithms}), and the role of the homepage feed on platforms (\Cref{sec:background:homepage}).

\subsection{Platform affordances} \label{sec:background:affordances}
The concept of affordances originated in ecological psychology to capture the relationship between an environment and its inhabitants \cite{Gibson1983TheSystems}. Researchers have widely adopted this concept in design and human-computer interaction studies \cite{Norman1988TheThings.}.
In the context of social media, affordances can be thought of as the ``{\em technological capabilities that provide the potential for a particular action}'' \cite{Majchrzak2013TheSharing, Faraj2012ThePerspective}. This definition, adopted in prior social computing literature \cite{Devito2017PlatformsMedia, Ellison2015SocialProcesses}, helps identify high-level affordances applicable to social media sites broadly (\eg searchability \cite{boyd2010SocialImplications}, feedback \cite{Devito2017PlatformsMedia}, and association \cite{Treem2013SocialAssociation}).
The affordance approach has been used to identify common patterns across platforms and to contextualize research findings in higher-level platform characteristics, rather than platform-specific details \cite{Devito2017PlatformsMedia, Theocharis2023PlatformEngagement, Vitak2014YouProcess}.
On algorithm-driven platforms, the affordance approach also serves to capture the mental constructs that people use to make sense of algorithmic outputs, which in turn shapes how they perceive the affordances of the platform \cite{Mayworm2024ContentUsers, Devito2018HowSelf-presentation, Devito2021AdaptivePlatforms, Caplan2020TieredEconomy}.
We use the concept of affordances to identify specific engagement mechanics on \yt, \X, and Reddit for our experiments and to compare affordances across these platforms. 
Due to high infrastructure costs (\cf \Cref{sec:discussion:limitations}), we focused on evaluating the influence of only the most popular mechanisms for exercising the following affordances, common to \yt, \X, and Reddit.

\para{Search.} Originally defined by boyd \cite{boyd2010SocialImplications} as one of four affordances that emerge in networked publics, the {\em search affordance is the extent to which a platform facilitates locating content}. 
Each platform in our study provided only a single mechanism for exercising this affordance --- a search bar at the top of each page. We used the default search settings in our experiments to search for content and altered the search configuration when the experiment needed to search for users or channels.

\para{Association.} Defined slightly differently in prior works \cite{Devito2017PlatformsMedia, ORiordan2012ExploringNetworks, Treem2013SocialAssociation}, the {\em association affordance broadly refers to the extent to which a platform facilitates linking of users to individuals or communities}. We refer to user-to-individual associations as U2I and user-to-community associations as U2C.
Each platform affords association via a variety of mechanisms.

\paraem{\yt.} Users can subscribe to creators (U2I) and join creator communities (U2C). Our experiments focused on U2I associations, as U2C associations require a monthly fee. Our experiments used the `Subscribe' button on the creator's homepage to exercise the U2I affordance.

\paraem{\X.} Users can follow others (U2I) and join communities (U2C). U2I mechanisms are available through homepage sidebars and user profile pages. U2C mechanisms are only available through the homepage sidebar. Our experiments examined the influence of the U2I profile page mechanism and the only U2C mechanism.

\paraem{Reddit.} Users can follow others (U2I) and join communities (U2C). U2I mechanisms are available through profile pages and pop-ups when hovering over a username. U2C associations can be made on subreddit homepages or by hovering over recommended content. We focused on U2C mechanisms on subreddit homepages in our experiments. We did not test the U2I mechanism as many users had disabled the `Follow' button on their profile.

\para{Feedback.} Defined in prior work \cite{ORiordan2012ExploringNetworks, Devito2017PlatformsMedia}, the {\em feedback affordance is the extent to which a platform affords direct responses to content}. 
Each platform in our study offers various feedback mechanisms, including sharing, commenting, liking/upvoting, and reporting. To manage the scope of our experiments, we focused on one positive feedback affordance for each platform: liking videos on \yt, liking tweets on \X, and upvoting posts on Reddit. Each feedback action was possible through a single mechanism: the \yt video page, the \X tweet body, and the Reddit post body.

\subsection{Content curation algorithms} \label{sec:background:algorithms}
Content curation algorithms are the backbone of social media platforms, enabling them to deliver personalized content for their users.
These algorithms analyze user interactions and consumption behaviors to recommend content that enhances user engagement.
Here, we explain basic concepts in content curation and then provide an overview of the content curation algorithms used by \yt, \X, and Reddit.

\para{Collaborative filtering.}
Collaborative filtering \cite{Goldberg1992UsingTapestry}, a popular technique to predict users' affinity for specific pieces of content, is the first step in the recommendation algorithm for all platforms considered in this study.
The basic idea behind collaborative filtering is that the interests of one user can be predicted by the observed interests of other similar users.
Collaborative filtering can be broken down into two steps \cite{2024Recommendations:Developers}. 

\paraem{Identifying clusters of similar users and content.} 
Platforms construct user vectors using demographic (\eg location), network (\ie engagement with users), and behavior (\ie engagement with content) signals.
Similarly, content vectors are created using a variety of features including topic, user interactions, and others.
Documents from \X \cite{TheTwitterTeam2023TwittersAlgorithm} and Facebook \cite{FacebookPapers, Narayanan2023UnderstandingAlgorithms} suggest that the engagement scores in these vectors are calculated by assigning weights to specific interactions (\eg likes, shares) and using weighted sums\footnote{It should be noted that these weights do not directly relate to the reach of the content (\ie content with 2x the engagement score is not likely to have twice the reach), rather they are indicative of the likelihood that the content will spur engagement from the specific user or cluster. The reach of content depends on many other factors, including considerations made by ranking algorithms.}.
%
%{\em In this paper, we use experimental data to derive the approximate values of these weights. \note{not sure if this is correct. need to think it through.}}
%
In the context of collaborative filtering, prior research has used a variety of metrics for computing similarities between these (user or content) vectors, including correlation-based metrics \cite{McLaughlin2004AExperience, Herlocker2004EvaluatingSystems}, conditional probability-based metrics \cite{Karypis2001EvaluationAlgorithms, Deshpande2004Item-basedAlgorithms}, and vector similarity metrics \cite{Sarwar2001Item-basedAlgorithms}.
These vectors and similarity measures are used for clustering to identify groups of users with similar behaviors and groups of content with similar characteristics. 
These clusters are used to identify candidates for recommendation (\eg content consumed by similar users or content similar to previously consumed content).

\paraem{Scoring recommendation candidates.}
The output from the prior step provides a matrix $M$ representing user feedback --- \ie $M_{ij}$ denotes user $i$'s engagement with content $j$. 
This matrix may have many unknown values --- \ie when user $i$ has never observed content $j$. The goal is to predict these unknown values. 
This may be achieved using:
(1) Matrix factorization where matrix factors which minimize error of known values in $M$ are identified \cite{Narayanan2023UnderstandingAlgorithms} and used to compute the unknown values in $M$; or 
(2) Training a deep neural network model \cite{2024Recommendations:Developers} which takes the user vector as input and generates an expected engagement vector as output.
These predicted values are the predicted engagement scores for recommendation candidates.

\paraem{Considerations when deploying collaborative filtering.} 
Implementing collaborative filtering-based recommender systems is not straightforward. 
Specifically, collaborative filtering deployments need to devise methods to address several challenges \cite{Shi2014CollaborativeMatrix, Terveen2001BeyondOther}. Relevant to this paper, this includes 
the {\em cold start problem} --- \ie how does the platform handle recommendations for users (or, content) with little/no interaction history.
Other challenges include large matrix factorization, sparse data, and handling users who do not cluster well with others.
%
%{\em In this paper, we also examine how the collaborative filtering algorithms of \yt, \X, and Reddit address these challenges (\Cref{sec:behaviors}).}

\para{The role of ranking algorithms.}
In simple cases, the predicted engagement scores from collaborative filtering can be used to create an ordering of content recommendations.
However, modern recommendation systems often use multiple sources of candidate recommendations and aim to satisfy multiple objectives beyond engagement.
For example, platforms may want to leverage the top-$k$ pieces of content from collaborative filtering, while exposing the user to trending or brand new content, promoting diversity, removing problematic content, and maximizing ad revenue.
In these cases, complex ranking algorithms are used to simultaneously satisfy these multiple objectives \cite{Zhao2023FairnessSurvey}.
The output of these ranking algorithms determines the ordering of the recommendation candidates. 
Top-ranked candidates are likely to be recommended on prime locations such as the top of the homepage feeds.

\para{Content curation algorithms deployed by social media platforms.}
We use publicly available documents from \yt, \X, and Reddit to provide an overview of their content curation systems.

\paraem{\yt.} 
Our understanding of the collaborative filtering and ranking systems used by \yt are obtained from  \yt blog posts \cite{Goodrow2021OnBlog, Meyerson2012YouTubeBlog} and a 2016 research article \cite{Covington2016DeepRecommendations}.
From these sources we see that \yt leverages two types of engagement features for collaborative filtering: {\em explicit features} afforded by the platform, such as liking, disliking, sharing, and others; and {\em implicit features} such as watch time and click-through rates.
Although the exact weights associated with each of these features is unclear, these documents indicate that \yt's collaborative filtering system skews more heavily towards utilizing implicit features, particularly watch times, because of the large amounts of implicit feature data available to them (likely because users are more likely to watch videos than explicitly engage with them).
Following collaborative filtering, the ranking algorithm, which aims to maximize engagement and fairness simultaneously, takes input from several sources, including content moderation algorithms to filter low-quality or problematic content, algorithms trained to identify authoritative content, and content from creators that a user is subscribed to (specifically for homepage feeds, not the ``Up Next'' panel). 

\paraem{\X.} 
Our understanding of the recommender system used by \X comes from the platform's source code (released in 2023 \cite{Twitter2023GitHubAlgorithm, Twitter2023The-algorithm-ml/projects/home/recapGitHub, Twitter2023The-algorithm-ml/projects/home/recap/FEATURES.mdGitHub}) and the accompanying blog post from the \X Engineering Team \cite{TheTwitterTeam2023TwittersAlgorithm}.
Unlike \yt, \X appears to rely almost exclusively on explicit engagement features (following, replying, liking, and others)
\footnote{We refer the reader to \cite{Twitter2023The-algorithm-ml/projects/home/recap/FEATURES.mdGitHub} for the comprehensive list of engagement features and \cite{Twitter2023The-algorithm-ml/projects/home/recapGitHub} for the weights associated with a small subset of these engagement features.}.
This makes sense given the nature of the platform and the low availability and fidelity of implicit engagement features. 
Interestingly, to ensure timeliness of recommended content the algorithm also associates expiration dates, ranging from 30 minutes to 50 days, with recorded engagements.
As part of the collaborative filtering process, \X uses a variant of matrix factorization \cite{Twitter2020GitHubTwitter/sbf} for mapping users and content into a embedding space and predicting users' engagement vector with content and other users. 
The \X homepage ranking algorithm then combines content from in-network sources and out-of-network sources at roughly equal proportions and aims to optimize author diversity and content quality.

\paraem{Reddit.}
We rely on Reddit blog posts \cite{/u/singmethesong2022TheR/reddit, /u/solutioneering2021EvolvingR/blog, /u/sacredtremor2021EvolvingR/RedditEng, Reddit2024RedditsHelp} to develop our understanding of Reddit's recommendation algorithms. 
We focus on the mechanics of the recommendation system used by the default homepage feed (sorted by ``Best'') and ignore other options (Hot, Top, New, and Rising).
Reddit, similar to \X, appears to rely exclusively on explicit engagement features. Also similar to \X, the collaborative filtering algorithm used by the Reddit homepage prioritizes timeliness of content. However, instead of associating engagement expiration dates, it does so by only scoring posts made within the past 24 hours. 
The Reddit algorithm uses a variety of demographic (location, device type, and account age), topic (inferred from history and recorded during on-boarding), community engagement (subscriptions, prior comments, views), and post engagement (votes, views, and comments) features for creating user vectors during collaborative filtering.
The algorithm then uses a deep neural network to predict whether the user will engage (view, vote, comment, or subscribe to source community) with the candidate content. These probabilities are fed to a scoring function, not publicly known, to produce an engagement score for each candidate post.
The ranking algorithm then orders the candidates using simple weighted sampling without other considerations.

\para{What remains unknown.}
The publicly available documents from \yt, \X, and Reddit provide many useful details about how recommendations are generated for homepage content. They shed light on the types of algorithms employed, the sources of data used for recommendation generation, considerations made during ranking, and (in the case of \X) even the weights associated with specific user interactions during collaborative filtering.
However, we still do not understand the effects of user interactions on the content curated for homepage feeds. Even when we know the weights and scoring functions for engagement, it is unclear how a specific interaction will influence a user’s subsequent recommendations. This uncertainty arises from the dynamics introduced by ranking algorithms, platform priorities, and the richness of their content libraries.
For example, engaging with a post about soccer should logically result in more soccer content being shown to the user. But how much more soccer content will appear? What other interests might the platform infer from this interaction? How does the algorithm’s behavior change in response? How can a user leverage platform affordances to influence the algorithm?
{\em We shed light on these questions by developing a methodology to systematically examine the effects of user interactions on homepage feeds.}

\subsection{Homepage feeds and platform behaviors} \label{sec:background:homepage}
Social media platforms offer many interfaces to observe the output of content curation algorithms, with the homepage feed being the most prominent. Previous research highlights the importance of the homepage feed in understanding how users consume \cite{Kumpel2019TheAll, Swart2021TacticsMedia, Tandoc2018AudiencesFramework, Guess2023HowCampaign} and control \cite{Hsu2020AwarenessOnline, Eslami2015IFeeds, Vaccaro2018TheSettings, Burrell2019WhenAlgorithms, Lu2020TamingPerspective} social media content.
Our study focuses on examining how engagement on a platform influences the content curated on their homepage feeds for two primary reasons. 
First, the homepage feed serves as the first interface when users sign in to their social media accounts and the primary interface through which users navigate the platform. 
Second, the importance of the homepage feed for user experience means that the content aggregated for it reflects the output of highly optimized content curation algorithms aimed to maximize user engagement and other platform priorities. 
In addition to using the homepage feed to understand how interactions influence recommendations, we also use to examine three specific behaviors of platforms: (1) their exploration-exploitation tendencies; (2) their reliance on explicit association signals; and (3) their dose-response behaviors.

\para{Exploration and exploitation tendencies.} The exploration-exploitation trade-off, originally introduced in the domain of organization management \cite{March1991ExplorationLearning}, reflects the dilemma faced when choosing between two opposing strategies: exploring new opportunities or exploiting existing opportunities. The trade-off arises because of limited resources and the inability to invest fully in both strategies.
In the context of social media platforms, recommender systems are faced with this dilemma when they curate content for a user. Should they rely on creating recommendations based on interests that they know the user to possess (exploitation) or should they make unexpected recommendations to add more knowledge about a user's interests (exploration)?
Previous work has explored algorithm designs to balance this trade-off  \cite{Barraza-Urbina2017TheSystems, Vanchinathan2014Explore-exploitProcesses, McInerney2018ExploreBandits} and examined their presence on TikTok \cite{Vombatkere2024TikTokFeeds} and Snapchat \cite{Gomez-Zara2024UnpackingData}.
Heavily exploitative behaviors may maximize immediate engagement, but result in diminishing user interest and problematic engagement patterns (\eg echo chambers and filter bubbles) in the long term. 
On the other hand, heavily exploratory behaviors may result in uncertain engagement behaviors due to the lack of personalized content, but yield a better understanding of user preferences in the long term.
In our study, {\em we uncover how platforms navigate the exploration-exploitation trade-off when curating users' homepage feeds}.

\para{Reliance on explicit association preference signals.} As shown in \Cref{sec:background:algorithms}, content curation algorithms use a variety of user engagement signals to make recommendations for their homepage feeds. 
Some of these signals explicitly indicate association preferences (\eg following a user, joining a community, and subscribing to a creator), while others are implicit indicators of interest that an algorithm must interpret (\eg viewing a video, liking a tweet, upvoting a post, or searching for a topic). 
Uncovering how platforms use these implicit and explicit signals is important for several reasons, including: (1) understanding the information propagation dynamics of platforms (\ie network-dominant vs. algorithm-dominant information propagation \cite{Narayanan2023UnderstandingAlgorithms, Goel2016TheDiffusion, Martin2016ExploringSystems}); and (2) understanding users' level of control over the content curation algorithms \cite{Hsu2020AwarenessOnline, Eslami2015IFeeds, Vaccaro2018TheSettings, Burrell2019WhenAlgorithms, Lu2020TamingPerspective}. 
Platforms that rely heavily on explicit association preferences will have a more predictable network-dominant propagation model, permitting users to have higher levels of control over their feeds and shared content.
In contrast, reliance on implicit preference signals leads to a more unpredictable algorithm-dominant propagation model, reducing user control and increasing the risk of context collapse \cite{Marwick2010IAudience}.
In our work, {\em we examine how platforms rely on explicit association preferences to curate their homepage feeds}.

\para{Dose-response behaviors.} The dose-response relationship, originally conceptualized in the context of pharmaceutical trials \cite{Ruberg1995DoseInterpretation}, describes the magnitude of the response of an organism to increasingly intense stimuli.
In the context of platform behaviors, the dose-response relationship highlights how a platform's algorithms respond to users whose interactions are homogeneous and repetitive (\ie increasing doses). Do platforms' responses further enable the homogeneous interactions by recommending more content related to the doses or do they mitigate these behaviors by forcing more diverse recommendations?
Interpreting the dose-response relationship can help understand the role of platforms in creating extremism, polarization, and filter bubbles, which may be downstream effects caused by such homogeneous interactions. These downstream effects have been the subject of much prior work \cite{Habib2022MakingMisogynist, Tucker2018SocialLiterature, Ribeiro2020AuditingYouTube, Tomlein2021AnChanges, Hussein2020MeasuringYouTube}.
In this work, {\em we examine how platforms' homepage feeds respond to multiple homogeneous interactions, occurring in succession, with a topic.}

% \note{Talk about the network vs engagement thing.}
% \note{Talk about why engagement optimization.}

% Across a variety of platforms that provide home feeds to its users, the composition of home feed is essentially determined by two sources: user subscriptions and user engagement profile. User subscriptions, as the name suggests, are sources the user is subscribed to. Analogous to a mailing list, content curated through this method is minimally algorithmic, ascribing greater control of home feed to the user. In contrast, content selected through the user interest profile is greatly reliant on algorithmic curation with the intent of curating relevant content that would be of interest to the user. To this end, the platform needs to create a user interest profile—essentially building an understanding of the user’s interests. The underlying goal of this study is to understand how the home feed is curated for users through subscription and engagement signals.

\section{Experiment Design and Data Processing} \label{sec:methods}

In this section, we describe our experiment design (\Cref{sec:methods:experiment}), ethical considerations (\Cref{sec:methods:ethics}),  methods for measuring changes in homepage feed composition (\Cref{sec:methods:homepage}), and our calculations of observed average treatment effects (\Cref{sec:methods:treatmenteffect}).

\subsection{Experiment design} \label{sec:methods:experiment}
The goal of our experiment was to uncover the influence of specific engagement interactions on the composition of homepage feeds on \yt, \X, and Reddit. We accomplished this using a sock puppet audit \cite{Sandvig2014AuditingPlatforms} and a crossover trial design \cite{Bose2020CrossoverDesigns}.

\RestyleAlgo{ruled}
\SetKwComment{Comment}{/*}{*/}
\DontPrintSemicolon
\SetCommentSty{mycommfont}
\begin{algorithm}
\caption{Pseudocode of crossover trial experiment for a single platform.}
\label{alg:crossover}
\footnotesize
\KwIn{Topics ($T$), Interactions ($I$)}
%\KwResult{Homepage feed observations for all topics $T$ and all interactions $I$ from the crossover trial experiments.}

\SetKwFunction{FMain}{Main}
\SetKwProg{Fn}{Function}{ is}{end}
\Fn{\FMain{$T$, $I$}}
{
    \Comment*[l]{{\tt GenerateLatinSquares} returns a set of sequences of $T$ such that each topic occurs only once within each sequence and only once within each position.}
    {\it TopicSequences} $\gets$ {\tt GenerateLatinSquares}($T$)\;
    \ForEach{{Sequence} in {TopicSequences}}{
        \ForEach{{Interaction} in $I$}{
            %\Comment*[l]{We create four sockpuppets for each configuration.}
            \For{$i \gets 1$ \KwTo $4$}
            {
                {\it FeedObservations} $\gets$ {\tt RunSockpuppet}({\it Permutation}, {\it Interaction})\;
            }
        }
    }
    \Return {\it FeedObservations}
}
\vspace{1.5\baselineskip}
\SetKwFunction{FMain}{RunSockPuppet}
\SetKwProg{Fn}{Function}{ is}{end}
\Fn{\FMain{Sequence, Interaction}}
{
    {\tt Primer()}\;
    {\it FeedObservations}{\tt .append}({\tt GetHomepageFeed()})\;
    \ForEach{{Topic} in {Sequence}}{
        %\Comment{We perform the interaction five times for each topic (in order). The homepage feed is saved 20 minutes after each interaction.}
        \For{$i \gets 1$ \KwTo $5$}{
            {\tt PerformInteraction}({\it Interaction}, {\it Topic}, $i$)\;
            {\it FeedObservations}{\tt .append}({\tt GetHomepageFeed()})\;
        }
        {\tt Primer()}\;
    }
    \Return {\it FeedObservations}
}
\vspace{1.5\baselineskip}
\SetKwFunction{FMain}{Primer}
\SetKwProg{Fn}{Function}{ is}{end}
\Fn{\FMain{}}
{
    \Comment*[l]{Creates a common baseline by performing identical interactions for all sockpuppets and dampens any residual carryover effects from prior treatments.}
    \ForEach{$q$ in $[\text{``Breakfast Recipes'', ``Lunch Recipes'', ``Dinner Recipes''}]$}
    {
        \For{$i \gets 1$ \KwTo $5$}
        {
            {\tt PerformInteraction}({\it ``Like''}, $q$, $i$)\;
        }
    }    
    \Return{}
}
\vspace{1.5\baselineskip}
\SetKwFunction{FMain}{PerformInteraction}
\SetKwProg{Fn}{Function}{ is}{end}
\Fn{\FMain{Interaction, Topic, i}}
{
    \Switch{{\it Interaction}}
    {
        \Case{{\it ``Search''}}
        {
            \Comment{Search content for {\it Topic}}
            {\tt PerformContentSearch}({\it Topic})
        }
        \Case{{\it ``View''}}
        {
            \Comment{Search content for {\it Topic} and open the $i^{th}$ result.}
            {\tt PerformContentSearch}({\it Topic});~
            {\tt ViewContent}({$i$}) \;
        }
        \Case{{\it ``Like''}}
        {
            \Comment{Search content for {\it Topic}, open the $i^{th}$ result, and like the content.}
            {\tt PerformContentSearch}({\it Topic});~
            {\tt ViewContent}({$i$});~
            {\tt LikeCurrentContent}({})\;
            }
        \Case{{\it ``Join''}}
        {
            \Comment{Search communities for {\it Topic}, open the $i^{th}$ result, and join the community.}
            {\tt PerformCommunitySearch}({\it Topic});~
            {\tt ViewCommunity}({$i$});~
            {\tt JoinCurrentCommunity}({})\;
            }
        \Case{{\it ``Follow''}}
        {
            \Comment{Search users for {\it Topic}, open the $i^{th}$ result, and follow the user.}
            {\tt PerformUserSearch}({\it Topic});~
            {\tt ViewProfile}({$i$});~
            {\tt FollowCurrentUser}({})\;
            }
        \Case{{\it ``Control''}}
        {
            \Comment{Do nothing.}
            \Return{}
        }
    }
    {\tt Wait}({\it 20} minutes) \;
    \Return{}
}
\end{algorithm}

% \para{Crossover trial design.} In order to measure the influence of specific interactions on homepage feeds, we needed to untangle the effects of topic from our measurements. Not doing so could result in an incorrect understanding of interaction influences\footnote{For example, finding that the `Like' interaction accounts for 30\% of the changes in the homepage feed is different than finding that the `Like' interaction applied to posts on `Fitness' accounts for 30\% of the changes in the homepage feed.}.
% One way to achieve this is through a randomized controlled trial \cite{Hariton2018RandomisedResearch} where each sockpuppet is randomly assigned a topic with which it will perform an assigned interaction. Then, the average treatment effect of the assigned interaction can be identified because of the `averaging out' of the effects of the topic. Unfortunately, due to high experiment infrastructure costs associated with simultaneously operating hundreds of sockpuppets with unique accounts on each platform (\Cref{sec:discussion:limitations}), this option is not feasible.
% %
% One approach to circumvent this challenge is to use a crossover trial design. Here, we assign each sockpuppet to a sequence of topics with which it will perform an assigned interaction (\eg Search Topic 1, then Topic 3, then Topic 2) such that every sequence has a fixed number of sockpuppets. Then, because Topic $n$ occurred in every position of the sequence, the influence of the topic can be estimated (by averaging the effects observed after each application of Topic $n$). 

\para{Interactions and topics.} The sockpuppets in our experiments tested one of five types of interactions related to the platform affordances described in \Cref{sec:background:affordances}. These were: Search (search affordance), View, Like (positive feedback affordance), Follow (user-to-individual association affordance), and Join (user-to-community association affordance). 
Because interactions do not occur in isolation and need to be targeted at specific content or users, we selected three neutral and mainstream topics as the targets of our interaction --- ``Kansas City Chiefs'', ``Elections'', and ``Fitness''. These were selected because they were popular topics on all platforms, were mainstream at the time of the experiments, and would not result in the {\em sparse library effect} that could bias our analysis of platform behaviors\footnote{The sparse library effect occurs when algorithms do not have relevant content to recommend because of the sparsity of related content in their libraries.}.
Below, we describe how each sockpuppet interaction was instrumented.

\paraem{Search interactions.} In order to perform a search interaction, the sockpuppets used the search bar at the top of each platform's homepage. The query of the search executed by the sockpuppet was determined by the crossover trial design. No changes were made to the default settings of the search. To measure the dosage effect of search, the process was repeated five times. The homepage feed was recorded between each iteration.

\paraem{View interactions.} The sockpuppets were assigned topics whose content they were to view (based on the crossover trial design described below). The search bar was used to search for content related to the topic. The sockpuppets then clicked open the top unopened result returned by the platform. This process was repeated five times with the sockpuppet opening the $i^{th}$ search result on the $i^{th}$ iteration. The homepage feed was recorded between each iteration.

\paraem{Like interactions.} Sockpuppets performed the view interaction and then liked/upvoted the opened content using the like/upvote button attached to the content. Similar to the view interaction, the process was repeated five times, liking the top-five posts in order (one interaction per iteration). The homepage feed was recorded between each iteration.

\paraem{Join (U2C) interactions.} This interaction was only performed on \X and Reddit. \yt was ignored because joining a creator's community required a monthly fee. To join a community, the search bar was configured to return communities related to the input query. Sockpuppets then clicked open the top unopened community returned by the platform and then clicked the `Join' (\X) or `Subscribe' (Reddit) button. The process was repeated similar to our other interactions and the homepage was recorded between each iteration.

\paraem{Follow (U2I) interactions.} This interaction was only performed on \yt and \X. Reddit was ignored because many users on the platform had disabled the `Follow' button on their profile. The search bar was configured to return users (or creators) related to the input query. Sockpuppets clicked open the top unopened profile returned by the platform and then clicked the `Follow' (\X) or `Subscribe' (\yt) button. The process was repeated similar to our other interactions and the homepage was recorded between each iteration.

\para{Crossover trial design.} We use a (stratified, 3-period, 3-treatment, uniform) crossover uniform design to overcome the challenges and costs associated with creating and simultaneously controlling the large number of sockpuppets necessary for a statistically valid randomized controlled trial \cite{Hariton2018RandomisedResearch}.
The crossover trial design allows us to apply multiple treatments to the sockpuppet and obtain the average treatment effect of each treatment in a statistically efficient manner. For each platform, we set up the crossover trial as follows (\cf~ \Cref{alg:crossover} and \Cref{tab:crossover}).

\paraem{Assigning sockpuppets to treatment sequences.} In our experiments, we use an (Interaction, Topic) pair as a treatment. We begin by creating six groups (one for each treatment interaction and one control group which performs no interaction). Within each group, we create three subgroups, one for each sequence in the 3x3 Latin square generated from the list of topics \cite{STAT509DesignandAnalysisofClinicalTrials202415.3509}. The interaction and sequence assigned to the subgroup determines the order of treatments that sockpuppets in that subgroup will undergo. We assign four sockpuppets to each subgroup and the control group for each interaction (which is not assigned any sequence of treatments).
In total, we are able to test fifteen different treatments with only 80 sockpuppets for each platform. On each platform, this design yields a total of 12 data points from which the influence of each (interaction, topic) pair can be assessed, 36 data points from which the influence of each interaction can be assessed, and 80 data points from which the influence of each topic can be assessed. This is in contrast to a standard Randomized Controlled Trial \cite{Hariton2018RandomisedResearch} which requires one sockpuppet to obtain a single (interaction, topic) data point.

\paraem{Priming sockpuppets to account for cold-start effects.} \X and \yt have a cold-start effect --- they do not show any content on their homepages for users with no prior interaction history. This leads to a control profile with no content, making any measurements of treatment effects meaningless.
We accounted for this by using a set of `primer' interactions, to be used by all control and treatment sockpuppets when initialized, to create a baseline from which treatment effects could be be measured.
Specifically, we applied the `Like' interaction, as instrumented above, to three topics: ``Breakfast Recipes'', ``Lunch Recipes'', and ``Dinner Recipes''. 

\paraem{Accounting for carryover effects.} Although the crossover trial design allows for more statistical efficiency, it has the potential to introduce complications due to carryover effects. Carryover effects occur when the prior treatment influences the response of the current treatment. 
In general, these effects are only statistically influential when the carryover effects of one treatment are more influential than the carryover effects of another. When this is not the case, a uniform crossover design\footnote{A uniform crossover design is one where each treatment occurs only once in a sequence and only once in each position.}, is sufficient to ensure that the carryover effect does not confound the measured treatment effect.
In \Cref{sec:rq1:methods}, we show how we mitigate this and other confounding effects by: (1) leveraging a uniform crossover trial design; and (2) using our primer interactions between each treatment to {\it washout} prior treatment effects. 

\begin{table}[t]
    \centering
    \small
    \begin{tabular}{ccp{1.2in}p{1.2in}p{1.2in}}
        \toprule
        \textbf{Treatment} & & \multicolumn{3}{c}{\textbf{Treatment Topic Sequence}} \\
         \textbf{Interaction} & \textbf{Group} & \textit{Topic 1} & \textit{Topic 2} & \textit{Topic 3} \\
        \midrule
        \multirow{4}{*}{\textit{Search}} & \multirow{3}{*}{\textit{Treatment}} & Kansas City Chiefs & Elections          & Fitness \\
                                         &                                     & Elections          & Fitness            & Kansas City Chiefs \\
                                         &                                     & Fitness            & Kansas City Chiefs & Elections \\
                                         \cmidrule{2-5}
                                         & \textit{Control}                    & \multicolumn{3}{c}{-} \\
        \midrule
        \multirow{4}{*}{\textit{Open}} & \multirow{3}{*}{\textit{Treatment}} & Kansas City Chiefs & Elections          & Fitness \\
                                       &                                     & Elections          & Fitness            & Kansas City Chiefs \\
                                       &                                     & Fitness            & Kansas City Chiefs & Elections \\
                                       \cmidrule{2-5}
                                       & \textit{Control}                    & \multicolumn{3}{c}{-} \\
        \midrule
        \multirow{4}{*}{\textit{Like}} & \multirow{3}{*}{\textit{Treatment}} & Kansas City Chiefs & Elections          & Fitness \\
                                       &                                     & Elections          & Fitness            & Kansas City Chiefs \\
                                       &                                     & Fitness            & Kansas City Chiefs & Elections \\
                                       \cmidrule{2-5}
                                       & \textit{Control}                    & \multicolumn{3}{c}{-} \\
        \midrule
        \multirow{4}{*}{\textit{Join (U2C)}} & \multirow{3}{*}{\textit{Treatment}} & Kansas City Chiefs & Elections          & Fitness \\
                                       &                                     & Elections          & Fitness            & Kansas City Chiefs \\
                                       &                                     & Fitness            & Kansas City Chiefs & Elections \\
                                       \cmidrule{2-5}
                                       & \textit{Control}                    & \multicolumn{3}{c}{-} \\
        \midrule
        \multirow{4}{*}{\textit{Follow (U2I)}} & \multirow{3}{*}{\textit{Treatment}} & Kansas City Chiefs & Elections          & Fitness \\
                                         &                                     & Elections          & Fitness            & Kansas City Chiefs \\
                                         &                                     & Fitness            & Kansas City Chiefs & Elections \\
                                         \cmidrule{2-5}
                                         & \textit{Control}                    & \multicolumn{3}{c}{-} \\
        \bottomrule
    \end{tabular}
    \caption{{\bf Crossover trial design used by our study.} Four sockpuppets were allocated to each row. Sockpuppets in that row performed the treatment interaction on each topic in the order shown in the treatment topic sequence. Between each topic, the priming function was used to minimize carryover effects from the prior treatment interaction. Each treatment interaction had a corresponding control group which did not perform any interactions, except the priming functions.}
    \label{tab:crossover}
\end{table}

\para{Sockpuppet creation and automation.} Following the best practices for web measurements \cite{Jueckstock2021TowardsMeasurements, Ahmad2020ApophaniesWeb}, our sockpuppets were automated using a fully-fledged Chrome web browser (v124.0) driven by {\tt undetected-chromedriver} (v3.5.0) \cite{ultrafunkamsterdam2024GitHubIUAM} and Selenium (v4.9) \cite{Selenium2024Selenium}. Data collection was performed between 05/25/2024 and 06/02/2024.
We created one new account on each platform for each sockpuppet. Each account required us to supply a phone number and an email address. We utilized a Google Workspace to automate the creation of Google accounts with valid email addresses and used a phone number rental service to obtain phone numbers for each of our sockpuppets. 
The Google accounts created by our workspace each came with a clean \yt profile that we used for our sockpuppets. The account creation process on \X yielded many CAPTCHAs that required manual intervention and could not be fully automated. 

\subsection{Ethical considerations during data gathering} \label{sec:methods:ethics}
During the design of our experiments, we sought to minimize harm to platforms and their users. 

\para{Minimizing harm to platform users.} Our research required us to perform interactions with content, users, and communities on popular social media platforms. We took several steps to ensure that our interactions did not disrupt or distress platform users. 
First, we designed our interactions to minimize intrusiveness. Specifically, we decided not to create posts, share other users' posts, or send (public or private) messages to users. 
Second, we focused on mainstream topics as the targets of our interactions, ensuring that the top-five users, communities, or content results returned by the search functions on the platforms were already very popular. This allowed us to perform interactions with them while only causing negligible algorithmic effects (\eg amplification) for them and other users.
Finally, we did not put any users at risk by recording private information. We obeyed the principle of data minimization \cite{Shukla2022DataChallenges, Gelinas2021NavigatingCollection} by only gathering data necessary for our analysis (content and source of homepage feeds and search results). To further mitigate privacy risks, user names and group names were replaced with their hashes in our gathered data. Further, while we do plan to release our source code \footnote{https://osf.io/swtq4/?view\_only=471a63d0986c4a148fb6a5a64452100a}, we will not release our feed observation dataset.

\para{Minimizing harm to platforms.} Our research used automated agents to interact with online platforms. Unfortunately, this was necessary due to the absence of API access for data gathering. In fact, our work follows a long line platform audit research relying on automated agents \cite{Dunna-CSCW2022,Chandio2024HowSystems, Hussein2020MeasuringYouTube, Tomlein2021AnChanges, Duskin2024EchoSystem, Chen2015PeekingUber}. Sandvig \etal \cite{Sandvig2014AuditingPlatforms} categorize this type of audit as a `sock puppet audit' in their classification of platform audit techniques.
To minimize costs to the platform, we did not perform any large-scale or high-bandwidth measurements. Each platform only encountered a few thousand clicks over several weeks as a result of our testing and experiment deployment.
We note that our work is not a violation of the Computer Fraud and Abuse Act (CFAA) \cite{USDepartmentofJustice2022JusticeJustice} due to a recent Supreme Court ruling (Van Buren vs. United States \cite{SUPREMECOURTOFTHEUNITEDSTATES2020VANSTATES}). 
Overall, given the importance of our research questions and findings to  users and platforms, we strongly believe that our experiments respect the principle of beneficence outlined in the Menlo Report \cite{OfHomelandSecurity2012} and Belmont report \cite{USDepartmentofHealthandHumanServices2024TheHHS.gov}.

\RestyleAlgo{ruled}
\SetKwComment{Comment}{/*}{*/}
\DontPrintSemicolon
\SetCommentSty{mycommfont}
\begin{algorithm}
\caption{Pseudocode for creating homepage feed composition vectors.}
\label{alg:composition}
\footnotesize

\KwIn{Observed feeds ({\it Feeds}), Primer interactions ({\it Primers}), Treatment interactions ({\it Treatments})}
\SetKwFunction{FMain}{Main}
\SetKwProg{Fn}{Function}{ is}{end}
\SetKwFunction{GetEmbedding}{GetEmbedding}
\SetKwFunction{GetLabeledClusters}{GetLabeledClusters}
%\SetKwFunction{AssignClusterLabels}{AssignClusterLabels}
\SetKwFunction{AssignPostLabels}{AssignPostLabels}

\SetKwFunction{GetSourceType}{GetSourceType}
\SetKwFunction{GetPostRank}{GetPostRank}
\SetKwFunction{GetFeedLength}{GetFeedLength}
\SetKwFunction{GetTopicVector}{GetTopicVectors}
\SetKwFunction{GetSourceVector}{GetSourceVector}
\SetKwFunction{GetPostEmbedding}{GetPostEmbedding}
\SetKwFunction{GetCentroid}{ComputeMean}

\Fn{\FMain{{\it Feeds}, {\it Primers}, {\it Treatments}}}
{
    \Comment*[l]{Compute embeddings for each post.}
    \ForEach{{\it Feed} in {\it Feeds}}
    {
        \ForEach{{\it Post} in {\it Feed}}
        {
            {\it Embeddings}[{\it Post}] $\gets$ \GetEmbedding{Post}\;
        }
    }
    
    \Comment*[l]{Manually label clusters and assign label to all cluster members.}
    {\it LabeledClusters} $\gets$ \GetLabeledClusters{Embeddings}\;
    {\it LabeledPosts} $\gets$ \AssignPostLabels{LabeledClusters} \;

    \Comment*[l]{Compute vector representations of each homepage feed.}
    \ForEach{{\it Feed} in {\it Feeds}}
    {
        {\it TopicVectors}[{\it Feed}] $\gets$ \GetTopicVector{Feed, LabeledPosts}\;
        {\it SourceVectors}[{\it Feed}] $\gets$ \GetSourceVector{Feed, Treatments, Primers}\;

    }
    \Return{\it TopicVectors, SourceVectors}

}
\vspace{1.5\baselineskip}
\SetKwProg{Fn}{Function}{ is}{end}
\Fn{\GetTopicVector{Feed, LabeledPosts}}
{
    \ForEach{{\it Post} in {\it Feed}}
    {
        \Comment*[l]{Append the embedding of {\it Post} to the list of {\it FeedEmbeddings}.}
        {\it FeedEmbeddings}{\tt .append}(\GetPostEmbedding{Post})\;
        \Comment*[l]{Get the label and rank associated with the current {\it Post}.}
        {\it PostLabel} $\gets$ {\it LabeledPosts}[{\it Post}];
        {\it PostRank}$\gets$ \GetPostRank{Post, Feed}\;
        \Comment*[l]{Increment prevalence for {\it Post}'s topic label. All posts are equally weighted.}
        {\it PrevalenceVector}[{\it PostLabel}] += {1}/{\GetFeedLength{Feed}} \;
        \Comment*[l]{Increment prominence for {\it Post}'s topic label.}
        \Comment*[l]{Weights are inversely proportional to {\it Post}'s rank.}
        {\it ProminenceVector}[{\it PostLabel}] += {1}/({\it PostRank $\times$ \GetFeedLength{Feed}}) \;
    }
    \Comment*[l]{{\it GenTopicEmbedding} is the mean of all embeddings in the feed.}
    {\it GenTopicEmbedding} $\gets$  \GetCentroid{FeedEmbeddings}\;
    \Return{PrevalenceVector, ProminenceVector, GenTopicEmbedding}
}

\vspace{1.5\baselineskip}
\SetKwProg{Fn}{Function}{ is}{end}
\Fn{\GetSourceVector{Feed, Treatments, Primers}}
{
    \ForEach{{\it Post} in {\it Feed}}
    {
        \Comment*[l]{Label the source of {\it Post} as in-network or out-of-network .}
        \Comment*[l]{This is done based on the interactions in {\it Treatments} and {\it Primers}.} 
        {\it PostSourceType}$\gets$ \GetSourceType{Post, Treatments, Primers}\;
        \Comment*[l]{Get the rank associated with the current {\it Post}.}
        {\it PostRank}$\gets$ \GetPostRank{Post, Feed}\;
        \Comment*[l]{Increment prevalence for {\it Post}'s source type. All posts are equally weighted.}
        {\it PrevalenceVector}[{\it PostSourceType}] += {1}/{\GetFeedLength{Feed}}\;
        \Comment*[l]{Increment prominence for {\it Post}'s source type.}
        \Comment*[l]{Weights are inversely proportional to {\it Post}'s rank.}
        {\it ProminenceVector}[{\it PostSourceType}] += {1}/({\it PostRank $\times$ \GetFeedLength{Feed}}) \;
    }
    \Return{PrevalenceVector, ProminenceVector}
}
\end{algorithm}

\begin{table}[t]
    \centering
    \small
    \begin{tabular}{lp{4in}}
        \toprule
        \textbf{Notation} & \textbf{Definition} \\
        \midrule
        {\it TopicPrevalence(t, F)} & Fraction of the topic $t$ contained in the feed $F$. {\it [\Cref{eq:topicprevalence}]}\\
        {\it TopicProminence(t, F)} & Rank-weighted fraction of the topic $t$ contained in the feed $F$. {\it [\Cref{eq:topicprominence}]}\\
        {\it SourcePrevalence(s, F, H)} & Fraction of the source type $s$ contained in the feed $F$. Source types are `in-network' or `out-of-network' and calculated using interaction history $H$. {\it [\Cref{eq:sourceprevalence}]}\\
        {\it SourceProminence(s, F, H)} & Rank-weighted fraction of the source type $s$ contained in the feed $F$. Source types are `in-network' or `out-of-network' and calculated using interaction history $H$. {\it [\Cref{eq:sourceprominence}]}\\
        $AvgEmbedding(F)$ & Average embedding computed across all posts in feed $F$. Embeddings are obtained using Open AI's {\tt text-embedding-3-small} model. {\it [\Cref{eq:topicembedding}]}\\
        \midrule
        {\it Topic(p)}  & Denotes the topic category for the post $p$. Topic categories may be: NFL, Politics, Fitness, Cooking, or Other. {\it [\Cref{eq:topicprevalence,,eq:topicprominence}]}\\
        {\it Source(p, H)} & Denotes the source category for the post $p$ based on interaction history $H$. Source categories may be: in-network or out-of-network. {\it [\Cref{eq:sourceprevalence,,eq:sourceprominence}]}\\
        {\it Rank(p, F)}    & Denotes the position rank of the post $p$ in feed $F$. {\it [\Cref{eq:topicprominence,,eq:sourceprominence}]}\\
        \midrule
        $\Delta(v_{pre}, v_{post})$ & Denotes the difference between the post- and pre-treatment composition vectors $v_{pre}$ and $v_{post}$. {\it [\Cref{eq:delta_categories,,eq:delta_embeddings}]}\\
        $\Delta^{\text{group}(k)}_{t_i, a_j}$ & Denotes the change in homepage composition after applying the treatment topic $t_i$ and action $a_j$ for the $k^{th}$ sockpuppet in `group'. `group' may be the ($t_i$, $a_j$) treatment group or the control group corresponding to the ($t_i$, $a_j$) treatment. {\it [\Cref{eq:delta_categories,,eq:delta_embeddings,,eq:measured_effect}]}\\
        $\hat{\mu}_{t_i,a_j}$ & Refers to the observed treatment effect of treatment ($t_i$, $a_j$) from our experiments. It includes nuisance effects. {\it [\Cref{eq:measured_effect,,eq:nuisance}]}\\
        \bottomrule
    \end{tabular}
    \caption{Notation used in \Cref{sec:methods:homepage} along with their definitions and relevant equations.}
    \label{tab:methods:notation}
\end{table}

\subsection{Measuring changes in homepage feeds} \label{sec:methods:homepage}
Our study focuses on identifying changes to the homepage feeds that result from user engagement with the platform. 
To accomplish this we first converted homepages into quantifiable vector representations which capture their composition characteristics. Then, we used these representations to compute changes to the homepage feed composition. We explain these methods below. \Cref{tab:methods:notation} describes the notation used in this section.

\para{Creating homepage composition vectors.} To measure the composition characteristics of a homepage feed, we created a number of vectors as described below and illustrated in \Cref{alg:composition}.

\paraem{Topic category prevalence vectors}.
Our goal was to capture the {\em prevalence} of posts associated with the specific topics used in our treatments on each homepage feed recorded by our sockpuppets.
To accomplish this, we first converted the text of each homepage feed post\footnote{We used the \yt API to obtain full text titles and descriptions of \yt videos.} recorded in our study into a vector embedding representation using OpenAI's {\tt text-embedding-3-small} model \cite{OpenAIModel2024Jan}. 
Then, we performed $k$-means clustering on these vectors using the {\em elbow method} \cite{Thorndike1953WhoFamily} to determine the optimal $k$. This approach clustered our 386K unique posts\footnote{In total, we gathered 117K \yt videos, 129K \X posts, and 140K Reddit posts from sockpuppets' homepage feeds.} into 17 unique clusters (silhouette scores \cite{Rousseeuw1987Silhouettes:Analysis}: ranging between 0.39 and 0.46).
We manually assigned one of five topics (NFL, Politics, Fitness, Cooking, and Other) to each cluster after analyzing a random sample of twenty posts from them. Each post inherited the topic label assigned to its cluster.
These labels were chosen based on the topics used in our treatments (\cf \Cref{sec:methods:experiment}) and because they appropriately captured the content observed in each cluster.
Finally, for each homepage, we created a five-element vector denoting the fraction of posts with each label. 
More concretely, the topic prevalence vector ({\it TopicPrevalence}) for a given topic label ($t$) and feed ($F$) is computed as:
\begin{equation}
    {\it TopicPrevalence}({\it t, F}) = \sum_{{p \in F} \wedge {\textit{Topic}(p)} = {\it t}} \frac{1}{|F|} 
    \label{eq:topicprevalence}
\end{equation}
Here, {\it Topic(p)} returns the label assigned to the post $p$. These vectors capture the prevalence of posts related to NFL, Politics, Fitness, Cooking, and Other topics on each homepage feed.

\paraem{Topic category prominence vectors.}
Our previous vectors capture the prevalence of specific topics on homepage feeds but do not account the {\em prominence} of these topics --- \ie how they were prioritized by the platform's ranking algorithm. 
To account for this, we assigned a weight to each post so that higher-ranked posts are seen as more prominent than lower-ranked posts. These weights were generated using a Zipfian function (ranks are inversely proportional to the assigned weight).
For each homepage, we used these weights to create a five-element vector denoting the Zipf-weighted sum of posts with each label. 
More concretely, the topic prominence vector ({\it TopicProminence}) for a given topic label ($t$) and feed ($F$) is computed as:
\begin{equation}
    {\it TopicProminence}({\it t, F}) = \sum_{{p \in F} \wedge {\textit{Topic}(p)} = {\it t}} \frac{1}{{\it Rank(p, F)} \times |F|} 
    \label{eq:topicprominence}
\end{equation}
Here, {\it Topic(p)} returns the label assigned to the post $p$ and {\it Rank(p, F)} returns the rank of the post $p$ in feed $F$. These vectors capture the prominence of posts related to NFL, Politics, Fitness, Cooking and Other topics on each homepage feed.

\paraem{Average topic embeddings.}
Our topic category vectors only capture the distribution of topics across four categories (NFL, Politics, Fitness, and Cooking) and relegate any unrelated posts to the `Other' category.
They fail to capture changes in the topics of these `Other' posts that might result from our treatments. Therefore, we sought to create a vector representation of all the topics on a page. 
For each homepage feed in our dataset, we converted the text of each post into an embedding using OpenAI's {\tt text-embedding-3-small} model. We then computed the centroid of these embedding and used it as our average topic embedding. 
More concretely, the average topic embedding ({\it AvgEmbedding}) for a feed ($F$) is computed as:
\begin{equation}
    {\it AvgEmbedding}(F) = \frac{1}{|F|} \sum_{p \in F} E(p)
    \label{eq:topicembedding}
\end{equation}
Here, {\it E(p)} returns the embedding of the post $p$. This average topic embedding captures the `{\em average topic}' on a given homepage feed.

\paraem{Source category prevalence vectors.}
Our collection of topic vectors do not yield insights into whose content is being recommended by the homepage feed algorithms.
To account for this, we categorized each post seen on a homepage feed as `in-network' or `out-of-network'. 
Posts were categorized as `in-network' if their authors/sources were the subject of a prior treatment or priming interaction (\eg if the sockpuppet liked one of their posts during the priming or treatment interaction). All other sources were categorized as `out-of-network'.
Note that only the sources that were the subject of the priming function are considered as `in-network' for homepage feeds generated by the control sockpuppets.
Finally, for each homepage, we used these labels to create a two-element vector where each element denoted the proportion of in-network and out-of-network posts.
Similar to the topic prevalence vector, the source prevalence vector for a source type $s$, feed $F$, and sockpuppet interaction history $H$ can be defined as:
\begin{equation}
    {\it SourcePrevalence}({\it s, F, H}) = \sum_{{p \in F} \wedge {\textit{Source}(p, H)} = {\it s}} \frac{1}{|F|} 
    \label{eq:sourceprevalence}
\end{equation}
Here, {\it Source(p)} returns the source type (\ie in-network or out-of-network) for the post $p$, based on the interaction history $H$.

\paraem{Weighted source category prominence vectors.}
In order to account for prominence of sources, we once again assigned weights to each post using a Zipfian function. 
We then used these weights to create a two-element vector denoting the Zipf-weighted incidence rates of in-network and out-of-network sources. This was repeated for each homepage feed in our dataset.
Similar to before, only the sources that were the subject of the priming function are considered as `in-network' for homepage feeds generated by the control sockpuppets.
The source prominence vector for a source type $s$ and feed $F$ can be defined as:
\begin{equation}
    {\it SourceProminence}({\it s, F, H}) = \sum_{{p \in F} \wedge {{\it Source}(p, H)} = {\it s}} \frac{1}{{\it Rank(p, F)} \times |F|}
    \label{eq:sourceprominence}
\end{equation}
Here, {\it Source(p)} returns the source type (\ie in-network or out-of-network) of the post $p$, based on the interaction history $H$ and {\it Rank(p, F)} returns the rank of post $p$ in feed $F$.

\RestyleAlgo{ruled}
\SetKwComment{Comment}{/*}{*/}
\DontPrintSemicolon
\SetCommentSty{mycommfont}
\begin{algorithm}
\caption{Pseudocode for measuring changes in homepage composition.}
\label{alg:changes}
\footnotesize
\KwIn{Sockpuppet group ({\it Group}), Vector type ({\it Type}), Treatment topic ({\it Topic}), and \\Pre- and post-treatment vectors ($V_{pre}$, $V_{post}$)}
\SetKwFunction{FMain}{ComputeCompositionChange}
\SetKwFunction{GetEuclideanDistance}{GetEuclideanDistance}
\SetKwProg{Fn}{Function}{ is}{end}
\Fn{\FMain{Group, Type, Topic, $V_{pre}$, $V_{post}$}}
{    
    \Comment*[l]{For category vector inputs, compute the difference between their {\it Topic} components.} 
    \uIf{{\it Type} == {\it ``Category''}}
    {
    \Comment*[l]{If the sockpuppet associated with the vectors belongs to the control group we use the ``Cooking'' topic. Otherwise, we proceed with the treatment topic.} 
        \uIf{Group == {\it ``Control''}}
        {    
            {\it Topic} $\gets$ {\it ``Cooking''}\;
        }
        
        $\Delta \gets V_{post}[{\it Topic}] - V_{pre}[{\it Topic}]$ \;
    }
    \Comment*[l]{For embedding inputs, compute the Euclidean distance between them.} 
    \uElseIf{{\it Type} == {\it ``Embedding''}}
    {
        $\Delta \gets$ \GetEuclideanDistance{$V_{post}$, $V_{pre}$}\;
    }
    \Return{$\Delta$}
}
\end{algorithm}

\para{Quantifying changes in homepage composition.}
Using the methods outlined above, for each homepage feed in our dataset we created five vectors. These represented the average topic of the feed and the prevalence and prominence of topic categories and source categories.
We used the methods described below (and illustrated in \Cref{alg:changes}) to quantify the difference between the (pre-treatment, post-treatment) vectors for a sockpuppet. 
Note that for sockpuppets in the control group, no treatment was performed and the vectors represent their homepage feeds recorded at roughly the same time as their corresponding treatment sockpuppets.

\paraem{Category vectors.}
To compute the difference between any pair of (pre-treatment, post-treatment) topic category vectors, we computed the difference between the prevalence (or, prominence) values associated with treatment topic. 
For example, the difference between the `Politics' values were used when the `Election' topic was used in the treatment applied to the sockpuppet. 
For source category vectors, we computed the difference between the prevalence (or, prominence) values associated with in-network sources.
For sockpuppets in the control group, the difference in the `Cooking' values of the topic category vectors was used.
More concretely, the difference ($\Delta$) between a pair of (pre-treatment, post-treatment) category vectors ({\it vec}$_{pre}$, {\it vec}$_{post}$):
\begin{equation}  
    \Delta(vec_{pre}, vec_{post}) =
    \begin{cases}
    {\it vec}_{post}[{\it ``Cooking"}] - {\it vec}_{pre}[{\it ``Cooking"}] 
                            & {\it if~} {\it Group(vec)} = {\it ``Control"} \\
    {\it vec}_{post}[{\it Topic(vec)}] - {\it vec}_{pre}[{\it Topic(vec)}] 
     & {\it otherwise}
    \end{cases}
    \label{eq:delta_categories}
\end{equation}
Here, {\it Group(vec)} returns the group assigned to the sockpuppet that generated {\it vec}. This may be ``{\it Control}'' or some ({\it Topic}, {\it Interaction}) group. {\it Topic(vec)} returns the {\it Topic} associated with the sockpuppet's assigned group.

\paraem{Average topic embeddings.}
To measure the difference between a pair of (pre-treatment, post-treatment) average topic embeddings, we computed the Euclidean distance between the two embeddings. Therefore, the difference ($\Delta$) between a pair of (pre-treatment, post-treatment) average topic embeddings ({\it e}$_{pre}$, {\it e}$_{post}$) is represented as:
\begin{equation}
    \Delta(e_{pre}, e_{post}) = \sqrt{\sum_{i} ({\it e}_{post}[i] - {\it e}_{pre}[i])^2}
    \label{eq:delta_embeddings}
\end{equation}
We use the Euclidean distance rather than the Cosine similarity because we are interested in understanding the magnitude of changes caused by our treatment, rather than change in direction.

\subsection{Measuring observed treatment effects}\label{sec:methods:treatmenteffect}
The sockpuppets in each treatment group were paired with a set of control sockpuppets. We used these data to measure the effects associated with our treatments. This was done with a {\em difference-in-differences} approach, using only the homepage feeds at the end of the interactions discussed in \Cref{sec:methods:experiment}. 
Using the methods outlined above (\Cref{sec:methods:homepage}), we computed the changes in homepage composition for each treatment group ($\Delta^{\text{treatment}}$) and their corresponding control group ($\Delta^{\text{control}}$).
Then, we recorded the observed treatment effect, for a (topic $t_i$, interaction $a_j$) pair, as the average difference in changes observed by each pair of ($t_i, a_j$) treatment and corresponding control sockpuppets. More concretely:
\begin{align}
    \hat{\mu}_{t_i,a_j} & = \frac{1}{n^2}\sum_{k=1}^n\sum_{l=1}^n(\Delta^{\text{treatment}(k)}_{t_i, a_j} - \Delta^{\text{control}(k)}_{t_i, a_j})    
    \label{eq:measured_effect}
\end{align}
Here, $ \hat{\mu}_{t_i,a_j}$ refers to the observed treatment effect for the ($t_i$, $a_j$) treatment pair, $n$ refers to the number of sockpuppets which received the ($t_i, a_j$) treatment pair, $\Delta^{\text{treatment}(k)}_{t_i, a_j}$ denotes the change in homepage composition observed by the $k^{th}$ sockpuppet assigned to the ($t_i$, $a_j$) treatment, and $\Delta^{\text{control}(k)}_{t_i, a_j}$ refers to the change observed by the $k^{th}$ control sockpuppet for the ($t_i$, $a_j$) treatment. In our experiments, $n=4$.

\subsection{Takeaways}
We use a crossover trial design to gather data about the influence of five different interactions across three mainstream topics (\Cref{sec:methods:experiment}). This design provides the statistical efficiency needed to gather sufficient data to analyze the influence of each (topic, interaction) pair, topic, and interaction. 
For data gathering, we used automated agents to interact with platforms and record homepage feeds while seeking to minimize harm to platform users and platforms (\Cref{sec:methods:ethics}).
To facilitate analysis related to our research questions, we converted each recorded homepage into five different homepage composition vectors which capture the prevalence and prominence of each topic and source type (\Cref{sec:methods:homepage}). We then used a difference-in-differences approach, for each homepage composition vector, to compute the average observed treatment effect for each treatment (\Cref{sec:methods:treatmenteffect}).
Therefore, we now have measurements of average observed treatment effects for each (topic, interaction) and for each homepage composition vector type. In \Cref{sec:rq1}, we use this data to obtain approximations of the true treatment effects associated with each (topic, interaction) pair, each topic, and each interaction. In \Cref{sec:rq2}, we use this data to characterize platforms' response to user interactions.

\section{How do user interactions influence the topics shown in their homepage feeds?} \label{sec:rq1}

We now use our data to understand how user interactions influence  the topics in their homepage feeds. In \Cref{sec:rq1:methods}, we describe our methods for measuring the influence of specific treatment parameters on the measured treatment effects in \Cref{eq:measured_effect}. In \Cref{sec:rq1:results}, we show our results indicating the influence of specific interactions on the homepage composition, for each platform and interaction in our study.

\subsection{Methods for measuring the influence of treatment interactions} \label{sec:rq1:methods}
In \Cref{sec:methods:homepage}, we showed how to quantify the changes that occur in a sockpuppet's homepage feed between two points in time (pre-treatment and post-treatment). Now we focus on identifying the effects caused by various parameters of our treatments. We use two approaches: (1) computing average effects from our crossover trial design; and (2) modeling our observed effects as a constraint satisfaction problem and solving for the influence of various parameters. \Cref{tab:rq1:notation} describes the notation used in this section.
\begin{table}[t]
    \centering
    \small
    \begin{tabular}{p{.4in}p{4.8in}}
        \toprule
        \textbf{Notation} & \textbf{Definition} \\
        \midrule
        ${\mu}_{t_i,a_j}$ & Refers to the true treatment effect of treatment ($t_i$, $a_j$). {\it [\Cref{eq:nuisance})]}\\
        $\hat{\mu}_{t_i,a_j}$ & Refers to the observed treatment effect of treatment ($t_i$, $a_j$) from our experiments. It includes nuisance effects. {\it [\Cref{eq:measured_effect,,eq:nuisance}]}\\
        $\hat{\mu}_{t_i}$ & Refers to the observed treatment effect of treatment topic $t_i$ (across all actions). {\it [\Cref{eq:effects_topic})]}\\
        $\hat{\mu}_{a_j}$ & Refers to the observed treatment effect of treatment action $a_j$ (across all topics). {\it [\Cref{eq:effects_action})]}\\
        \midrule
        $T_k(t_i, a_j)$ & Refers to the treatment effect measured when the treatment ($t_i$, $a_j$) was applied at position $k$ in the crossover sequence. {\it [\Cref{eq:nuisance}]}\\
        $\lambda_{t_i, a_j}$ & Refers to the carryover effect that occurs after the application of ($t_i$, $a_j$) when the next treatment effect is measured. {\it [\Cref{eq:nuisance}]}\\
        $\lambda_w$ & Refers to the carryover effect that occurs after the washout interactions are applied when the next treatment effect is measured. {\it [\Cref{eq:nuisance}]}\\
        \midrule
        $f_1$, $f_2$, $f_3$ & Refers to the influence functions associated with topics, interactions, and account age, respectively. {\it [\Cref{eq:lp}]}\\
        \bottomrule
    \end{tabular}
    \caption{Notation used in \Cref{sec:rq1:methods} along with their definitions and relevant equations.}
    \label{tab:rq1:notation}
\end{table}

\para{Estimating average treatment effects from observed treatment effects.}
\Cref{tab:nuisance} shows the measured treatment effects after each treatment was applied (and effects seen in the corresponding controls are removed), along with nuisance variables $\lambda$ which denotes the crossover effect, $\rho$ which denotes the period effect, and $\gamma$ which denotes the sequence effect. 
From \Cref{tab:crossover}, we see that our average observed treatment effect across all occurrences of each (topic, interaction) pair in the crossover trial include nuisance effects. 
Put another way, our measurements of $\hat{\mu}_{t_i,a_j}$ (which is $\Delta^{\text{treatment}(k)}_{t_i, a_j} - \Delta^{\text{control}(k)}_{t_i, a_j}$) actually report:
    \begin{align}
    \small
        \hat{\mu}_{t_i,a_j} & = (1/3) \times \sum_k T_{k}(t_i,a_j) \\
                            & = \mu_{t_i,a_j} + (2/3)\times(\lambda_{t_{(i-1)}, a_j} - \lambda_w)
    \label{eq:nuisance}
    \end{align}
Here, $\mu_{t_i,a_j}$ denotes the true average treatment effect of the (topic $t_i$, interaction $a_j$) treatment, $\hat{\mu}_{t_i,a_j}$ denotes our estimate of the average treatment effect, and $T_k(t_i, a_j)$ denotes the treatment effect measured when the pair ($t_i$, $a_j$) occurred in the $k^{th}$ position of the treatment sequence. 
Our uniform crossover trial design allows us to remove all sequence effects and period effects (since each treatment occurs once in every sequence and once in every period \cite{Bose2020CrossoverDesigns}). Although our estimate is confounded by the carryover effect from neighboring topic $t_{i-1}$, we note that it is dampened by the carryover effect from our washout interactions ($\lambda_w$) which is observed in every measured treatment effect (making the treatment effects comparable with each other). Later in this section, we will show that this effect is marginal and statistically insignificant, making our recorded $\hat{\mu}_{t_i,a_j}$ (\Cref{eq:measured_effect}) reasonable estimates of the true treatment effect of a (topic, interaction) pair. We also report our estimates for the average effects $\hat{\mu}_{t_i}$ and $\hat{\mu}_{a_j}$ of each topic ($t_i$) and interaction ($a_j$) as follows:
    \begin{align}
        \hat{\mu}_{t_i} & = \frac{1}{n} \sum_{j=1}^{n} \hat{\mu}_{t_i, a_j}\label{eq:effects_topic}\\
        \hat{\mu}_{a_j} & = \frac{1}{m} \sum_{i=1}^{m} \hat{\mu}_{t_i, a_j}
    \label{eq:effects_action}
    \end{align}

\paraem{Testing significance and magnitude of carryover effects.}
As shown above, our estimates ($\hat{\mu}_{t_i, a_j}$) do include a carryover effect of $(2/3) \times (\lambda_{t_(i-1), a_j} - \lambda_w$).
To test whether this carryover effect was significant, we followed a standard approach \cite{Bose2020CrossoverDesigns, Shen2006EstimateDesigns}, using an ANOVA to test for the differences in the sums of treatment effects observed by sockpuppets in each treatment sequence, within and across all interactions --- \ie we tested $H^{a_j}_0: \sum_i M^{a_j}_{1,i} = \sum_i M^{a_j}_{2,i} = \sum_i M^{a_j}_{3,i}$, where $M^{a_j}_{s, i}$ denotes the value of the treatment effect measured at $T_j$ when the sequence $s$ is applied with interaction $a_j$. This effectively tests for a statistical difference in the average carryover effects after each treatment sequence is applied. We observed no statistically significant differences in their means, for all but one case. Ignoring the significance of the results, even the magnitudes of the differences in means were small (mean differences ranged from .03\% to 4.6\%).
Only on Reddit, the {\it Open} interaction was found to have a statistically significant difference, albeit of marginal magnitude (mean difference of 1.2\%). Therefore, with the exception of the Open interaction on Reddit, the carryover effects can be ignored and we can use our measured treatment effects as reasonable approximations of the actual treatment effect.

\begin{table}[t]
    \footnotesize
    \begin{tabular}{ccccc}
        \toprule
        \multirow{1}{*}{{\bf Interaction}} & {\bf Topic Sequence} &  {$T_{1}$} & ${T_{2}}$ & $T_{3}$ \\ 
        \midrule
        \multirow{3}{*}
        {{\it Search (S)}}        & $seq_1$: (AwBwC) & $\mu_{AS} + \rho_{1} + \gamma_{1} $ & $\mu_{BS} + \rho_{2} + \gamma_{1} - (\lambda_{w} - \lambda_{AS})$ & $\mu_{CS} + \rho_{3} + \gamma_{1} - (\lambda_{w} - \lambda_{BS})$\\\cline{2-5}
                                  & $seq_2$: (BwCwA) & $\mu_{BS} + \rho_{1} + \gamma_{2} $ & $\mu_{CS} + \rho_{2} + \gamma_{2} - (\lambda_{w} - \lambda_{BS})$ & $\mu_{AS} + \rho_{3} + \gamma_{2} - (\lambda_{w} - \lambda_{CS})$\\\cline{2-5}
                                  & $seq_3$: (CwAwB) & $\mu_{CS} + \rho_{1} + \gamma_{3} $ & $\mu_{AS} + \rho_{2} + \gamma_{3} - (\lambda_{w} - \lambda_{CS})$ & $\mu_{BS} + \rho_{3} + \gamma_{3} - (\lambda_{w} - \lambda_{AS})$\\ \midrule
        \multirow{3}{*}
        {{\it Open (O)}}          & $seq_1$: (AwBwC) & $\mu_{AO} + \rho_{1} + \gamma_{1} $ & $\mu_{BO} + \rho_{2} + \gamma_{1} - (\lambda_{w} - \lambda_{AO})$ & $\mu_{CO} + \rho_{3} + \gamma_{1} - (\lambda_{w} - \lambda_{BO})$\\\cline{2-5}
                                  & $seq_2$: (BwCwA) & $\mu_{BO} + \rho_{1} + \gamma_{2} $ & $\mu_{CO} + \rho_{2} + \gamma_{2} - (\lambda_{w} - \lambda_{BO})$ & $\mu_{AO} + \rho_{3} + \gamma_{2} - (\lambda_{w} - \lambda_{CO})$\\\cline{2-5}
                                  & $seq_3$: (CwAwB) & $\mu_{CO} + \rho_{1} + \gamma_{3} $ & $\mu_{AO} + \rho_{2} + \gamma_{3} - (\lambda_{w} - \lambda_{CO})$ & $\mu_{BO} + \rho_{3} + \gamma_{3} - (\lambda_{w} - \lambda_{AO})$\\ \midrule
        \multirow{3}{*}
        {{\it Like (L)}}          & $seq_1$: (AwBwC) & $\mu_{AL} + \rho_{1} + \gamma_{1} $ & $\mu_{BL} + \rho_{2} + \gamma_{1} - (\lambda_{w} - \lambda_{AL})$ & $\mu_{CL} + \rho_{3} + \gamma_{1} - (\lambda_{w} - \lambda_{BL})$\\\cline{2-5}
                                  & $seq_2$: (BwCwA) & $\mu_{BL} + \rho_{1} + \gamma_{2} $ & $\mu_{CL} + \rho_{2} + \gamma_{2} - (\lambda_{w} - \lambda_{BL})$ & $\mu_{AL} + \rho_{3} + \gamma_{2} - (\lambda_{w} - \lambda_{CL})$\\\cline{2-5}
                                  & $seq_3$: (CwAwB) & $\mu_{CL} + \rho_{1} + \gamma_{3} $ & $\mu_{AL} + \rho_{2} + \gamma_{3} - (\lambda_{w} - \lambda_{CL})$ & $\mu_{BL} + \rho_{3} + \gamma_{3} - (\lambda_{w} - \lambda_{AL})$\\ \midrule
        \multirow{3}{*}
        {{\it (U2C) Join (J)}}          & $seq_1$: (AwBwC) & $\mu_{AJ} + \rho_{1} + \gamma_{1} $ & $\mu_{BJ} + \rho_{2} + \gamma_{1} - (\lambda_{w} - \lambda_{AJ})$ & $\mu_{CJ} + \rho_{3} + \gamma_{1} - (\lambda_{w} - \lambda_{BJ})$\\\cline{2-5}
                                  & $seq_2$: (BwCwA) & $\mu_{BJ} + \rho_{1} + \gamma_{2} $ & $\mu_{CJ} + \rho_{2} + \gamma_{2} - (\lambda_{w} - \lambda_{BJ})$ & $\mu_{AJ} + \rho_{3} + \gamma_{2} - (\lambda_{w} - \lambda_{CJ})$\\\cline{2-5}
                                  & $seq_3$: (CwAwB) & $\mu_{CJ} + \rho_{1} + \gamma_{3} $ & $\mu_{AJ} + \rho_{2} + \gamma_{3} - (\lambda_{w} - \lambda_{CJ})$ & $\mu_{BJ} + \rho_{3} + \gamma_{3} - (\lambda_{w} - \lambda_{AJ})$\\ \midrule
        \multirow{3}{*}
        {{\it (U2I) Follow (F)}}        & $seq_1$: (AwBwC) & $\mu_{AF} + \rho_{1} + \gamma_{1} $ & $\mu_{BF} + \rho_{2} + \gamma_{1} - (\lambda_{w} - \lambda_{AF})$ & $\mu_{CF} + \rho_{3} + \gamma_{1} - (\lambda_{w} - \lambda_{BF})$\\\cline{2-5}
                                  & $seq_2$: (BwCwA) & $\mu_{BF} + \rho_{1} + \gamma_{2} $ & $\mu_{CF} + \rho_{2} + \gamma_{2} - (\lambda_{w} - \lambda_{BF})$ & $\mu_{AF} + \rho_{3} + \gamma_{2} - (\lambda_{w} - \lambda_{CF})$\\\cline{2-5}
                                  & $seq_3$: (CwAwB) & $\mu_{CF} + \rho_{1} + \gamma_{3} $ & $\mu_{AF} + \rho_{2} + \gamma_{3} - (\lambda_{w} - \lambda_{CF})$ & $\mu_{BF} + \rho_{3} + \gamma_{3} - (\lambda_{w} - \lambda_{AF})$\\\bottomrule                                  
    \end{tabular}
    \caption{\textbf{Nuisance effects observed in the measured treatment effects.} Four sockpuppets were allocated to each row. Each cell denotes the measured effect size with nuisance variables.
    The topics A, B, C, and w refer to {\it Kansas City Chiefs}, {\it Elections}, {\it Fitness}, and {\it Cooking} (washout and primer topic). 
    $T_1$, $T_2$, and $T_3$ refer to the treatment effects measured after the application of Topic 1, Topic 2, and Topic 3, respectively. $seq_i$ refers to the $i^{th}$ Latin square sequence.
    $\mu_{ij}$ refers to the effect of applying interaction $j$ on topic $i$. 
    $\lambda_{ij}$ refers to the carryover effect from applying interaction $j$ on topic $i$. $\lambda_{w}$ refers to the carryover effect of our washout interactions. $\rho_{k}$ refers to the period effect at the time of treatment $k$. $\gamma_{k}$ refers to the sequence effect for the $k^{th}$ topic sequence. Note that $\sum_i \rho_i = 0$ and $\sum_i \gamma_i = 0$. $M_{i, j}$ denoted the value of the measured treatment effect for $seq_i$ at $T_j$.}
    \label{tab:nuisance}
\end{table}

\para{Modeling as a constraint satisfaction problem (CSP) to isolate influence of treatments.}
Each of our experiments provides us with an effect size that can be attributed to a combination of three factors: the interaction being applied, the topic that is the subject of the interaction, and the account age (which also includes carryover effects with prior interactions).
Put differently, each of our observed treatment effects $\hat{\mu}_{t_i,a_j}$ recorded from sequence position $seq_k$ can be viewed as being the product of three functions: $f_1(t_i)$, $f_2(a_j)$, and $f_3(seq_k)$. Here, $f_1$ represents the topic influence function, $f_2$ represents the interaction influence function, and $f_3$ represents the account age and carryover influence function.
This formulation allows us to organize our observations as a series of linear equations as follows. 
\begin{align}
    \log(\hat{\mu}^{(seq_k)}_{t_i,a_j}) & = \log(f_1(t_i)) + \log(f_2(a_j)) + \log(f_3(seq_k)) & \forall i; \forall j; \forall k 
    \label{eq:lp}
\end{align}
We then solve this set of equations, obtaining the influence for each $\log(f_1(t_i))$, $\log(f_2(a_j))$, and $\log(f_3(seq_k))$, with linear programming. We recover the values associated with $f_1(t_i)$, $f_2(a_j)$, and $f_3(seq_k)$ by using the $\exp()$ function. Note that the corresponding co-efficient matrix is square and has full rank, indicating that our solution is unique \cite{Strang2012LinearApplications}. Values indicate the magnitude of the influence associated with the specific $t_i$, $a_j$, or $seq_k$. %Influence values larger than 1 are interpreted as having an amplifying effect and values smaller than 1 are interpreted as having a dampening effect.

\begin{table}[h!]
    \centering
    \small
    \begin{adjustbox}{max width=\textwidth}
    \begin{tabular}{cclcccccc}
        \toprule
        & & & & \multicolumn{5}{c}{\textbf{Actions}} \\
        \cmidrule(lr){5-9}
        \textbf{Platform} & \textbf{Topic} & \textbf{Measure} & \textbf{$\hat{\mu}_{topic}$} & \textit{Search} & \textit{Open} & \textit{Like} & \textit{Join (U2C)} & \textit{Follow (U2I)} \\
        \midrule
        \multirow{12}{*}{YouTube} 
            & \multirow{3}{*}{NFL}      & {\it TopicPrevalence} & \textcolor{red}{10.39*} & 0.71 & \textcolor{red}{20.05*} & \textcolor{red}{22.61*} & - & \textcolor{red}{5.13*} \\
            &                           & {\it TopicProminence} & \textcolor{red}{14.52*} & -0.3 & \textcolor{red}{26.8*} & \textcolor{red}{29.8*} & - & \textcolor{red}{9.8*} \\
            % ?? & ?? & ?? & -  & ??  \\
            &                           & {\it AvgEmbedding} &  0.91  & 0.29 & \textcolor{red}{1.6*} & \textcolor{red}{1.50*} & - & 0.23 \\
        \cmidrule(lr){2-9}
            & \multirow{3}{*}{Politics} & {\it TopicPrevalence} & \textcolor{red}{13.31*} & 2.65 & \textcolor{red}{20.98*} & \textcolor{red}{34.39*} & - & \textcolor{red}{6.4*} \\
            &                           & {\it TopicProminence} &  \textcolor{red}{18.08*} &  2.9 & \textcolor{red}{27.6*} & \textcolor{red}{42.6*} & -  & \textcolor{red}{19.0*} \\
            &                           & {\it AvgEmbedding} &  0.89 & -0.0 & \textcolor{red}{1.11*} & \textcolor{red}{2.3*} & - & 0.16 \\
        \cmidrule(lr){2-9}
            & \multirow{3}{*}{Fitness}  & {\it TopicPrevalence} & \textcolor{red}{7.57*} & 0.9 & \textcolor{red}{14.95*} & \textcolor{red}{13.78*} & - & 4.51 \\
            &                           & {\it TopicProminence} &  \textcolor{red}{10.78*} &  -1.5 & \textcolor{red}{16.6*} & \textcolor{red}{18.2*} & - & \textcolor{red}{11.1*} \\
            &                           & {\it AvgEmbedding} &  0.25 & 0.24 & \textcolor{red}{0.47*} & \textcolor{red}{0.69*} & - & 0.42 \\
        \cmidrule(lr){2-9}
            & \multirow{3}{*}{\textbf{$\hat\mu_{action}$}}  & {\it TopicPrevalence} &  - & 0.43 & \textcolor{red}{17.49*} & \textcolor{red}{21.25*} & - & \textcolor{red}{4.34*} \\
            &                                               & {\it TopicProminence} & - & 0.04 & \textcolor{red}{23.25*} & \textcolor{red}{28.61*} & - & \textcolor{red}{12.62*} \\
            &                                               & {\it AvgEmbedding} & - & 0.17 & \textcolor{red}{1.06*} & \textcolor{red}{1.42*} & - & 0.28 \\
        \bottomrule
        \multirow{12}{*}{\X} 
            & \multirow{3}{*}{NFL}      & {\it TopicPrevalence} & \textcolor{red}{1.95*} & 0.02 & 0.7 & 3.91 & 0.41 & \textcolor{red}{6.87*} \\
            &                           & {\it TopicProminence} &  8.54 &  -2.8 &  -3.3 & 10.8  & -0.2 &  \textcolor{red}{12.2*} \\
            &                           & {\it AvgEmbedding} &  0.30 & 0.10 & -0.0 & \textcolor{red}{0.45*} & 0.23 & \textcolor{red}{0.74*} \\
        \cmidrule(lr){2-9}
            & \multirow{3}{*}{Politics} & {\it TopicPrevalence} & -0.80 & -1.91 & 0.15 & -2.32 & -0.12 & 2.31 \\
            &                           & {\it TopicProminence} &  -3.73 & -4.8 & 2.74 & -5.4 & -0.2  & 1.71  \\
            &                           & {\it AvgEmbedding} &  0.19 & 0.06 & -0.0 & 0.44 & 0.18 & \textcolor{red}{0.3*} \\
        \cmidrule(lr){2-9}
            & \multirow{3}{*}{Fitness}  & {\it TopicPrevalence} & 2.04 & 0.42 & 1.74 & \textcolor{red}{6.03*} & 1.75 & 3.54 \\
            &                           & {\it TopicProminence} &  4.13 & -1.2 & -0.2 & \textcolor{red}{13.0*} & 2.68 & 0.79  \\
            &                           & {\it AvgEmbedding} & 0.06 & -0.1 & 0.13 & 0.27 & 0.00 & 0.0 \\
        \cmidrule(lr){2-9}
            & \multirow{3}{*}{\textbf{$\hat\mu_{action}$}}  & {\it TopicPrevalence} & - & -0.83 & 0.6 & 2.72 & 0.36 & \textcolor{red}{2.6*} \\
            &                                               & {\it TopicProminence} & - & -3.2 & -0.72 & \textcolor{red}{7.44*} & 0.73 & \textcolor{red}{5.68*} \\
            &                                               & {\it AvgEmbedding} & - & 0.0 & 0.03 & \textcolor{red}{0.39*} & 0.13 & 0.22 \\
        \midrule
        \multirow{12}{*}{Reddit} 
            & \multirow{3}{*}{NFL}      & {\it TopicPrevalence} & \textcolor{red}{11.53*} & 1.98 & \textcolor{red}{10.75*} & \textcolor{red}{10.59*} & \textcolor{red}{28.37*} & - \\
            &                           & {\it TopicProminence} &  \textcolor{red}{13.34*} & 0.18  & \textcolor{red}{9.03*} & \textcolor{red}{11.6*} & \textcolor{red}{33.1*} & -  \\
            &                           & {\it AvgEmbedding} & 0.56 & 0.0 & \textcolor{red}{0.33*} & 0.26 & \textcolor{red}{1.64*} & - \\
        \cmidrule(lr){2-9}
            & \multirow{3}{*}{Politics} & {\it TopicPrevalence} & 6.98 & 0.86 & \textcolor{red}{3.63*} & -0.33 & \textcolor{red}{28.35*} & - \\
            &                           & {\it TopicProminence} &  5.78 & 0.18 & \textcolor{red}{1.56*} & -3.7 & \textcolor{red}{29.1*} & -  \\
            &                           & {\it AvgEmbedding} & 0.66 & 0.03 & 0.09 & \textcolor{red}{0.18*} & \textcolor{red}{2.33*} & - \\
        \cmidrule(lr){2-9}
            & \multirow{3}{*}{Fitness} & {\it TopicPrevalence} & 2.23 & 0.27 & 2.21 & -0.04 & \textcolor{red}{12.29*} & - \\
            &                           & {\it TopicProminence} &  3.92 & 0.03 & 0.0 & 0.0  & \textcolor{red}{18.5*} & -  \\
            &                          & {\it AvgEmbedding} & 0.30 & 0.0 & 0.13 & 0.0 & \textcolor{red}{1.06*} & - \\
        \cmidrule(lr){2-9}
            & \multirow{3}{*}{\textbf{$\hat\mu_{action}$}}  & {\it TopicPrevalence} & - & 0.36 & \textcolor{red}{4.22*} & \textcolor{red}{3.26*} & \textcolor{red}{21.72*} & - \\
            &                                               & {\it TopicProminence} & - & 0.12 & \textcolor{red}{3.74*} & 3.57 & \textcolor{red}{26.83*} & - \\
            &                                               & {\it AvgEmbedding} & - &  0.0 & \textcolor{red}{0.19*} & 0.14 & 1.59* & - \\
        \bottomrule
    \end{tabular}
    \end{adjustbox}
    \caption{Influence of different (topic, action) pairs on the topic composition vectors of the homepage feed on \yt, \X, and Reddit. Each cell indicates the measured average $\hat{\mu}_{\text{topic, action}}$ treatment effect. $\hat{\mu}_{\text{topic}}$ and $\hat{\mu}_{\text{action}}$ denote the average treatment effect for each topic (across all actions) and action (across all topics), respectively. Values marked with \textcolor{red}{red*} are statistically different from the control feeds ($p < .05$).}
    \label{tab:topic-results}
\end{table}

\begin{table}[t]
\centering
\small
\resizebox{\textwidth}{!}{%
\begin{tabular}{@{}lccccccccccc@{}}
\toprule
\textbf{} & \multicolumn{5}{c}{\textbf{Actions}} & \multicolumn{3}{c}{\textbf{Topics}} & \multicolumn{3}{c}{\textbf{Treatment Position}} \\ 
\cmidrule(lr){2-6} \cmidrule(lr){7-9} \cmidrule(lr){10-12}
 \textbf{Platform} & \textit{Search} & \textit{Open} & \textit{Like} & \textit{Join (U2C)} & \textit{Follow ({U2I})} & \textit{NFL} & \textit{Politics} & \textit{Fitness} & $Seq_1$ & $Seq_2$ & $Seq_3$ \\ 
\midrule
YouTube  & 0.50  & 1.87  & \textcolor{red}{36.52}  & -  & 3.00  & 3.70  & \textcolor{red}{6.19}  & 0.80  & \textcolor{red}{3.79}  & 2.70  & 1.79  \\
\X  & 0.64  & 0.76  & \textcolor{red}{2.16}   & 0.83  & 1.54  & 1.19  & 0.83  & \textcolor{red}{1.23}  & \textcolor{red}{1.43}  & 0.93  & 0.90  \\ 
Reddit   & 0.23  & 0.72  & 0.85   & \textcolor{red}{301.06}  & -  & \textcolor{red}{6.77}  & 0.78  & 1.95  & \textcolor{red}{4.33}  & 2.05  & 1.17  \\
\bottomrule
\end{tabular}%
}
\caption{Values assigned to influence variables from the constraint satisfaction problem formulated using topic prevalence vectors. The most influential action, topic, and position are highlighted in \textcolor{red}{red}.}
\label{tab:constraint-topicprevalence}
\end{table}

\subsection{Results} \label{sec:rq1:results}
We now present the results of our analysis as described in \Cref{sec:rq1:methods}. They are summarized in \Cref{tab:topic-results,,tab:constraint-topicprevalence}. We breakdown our results by platform and then contrast their behaviors in our final takeaway from this analysis. 

\para{\yt.} Our data provides several interesting insights into \yt's homepage feed algorithms. 

\paraem{High variance of action influence.}
We see a high variance in measured treatment effects for each action indicating that the platform relies heavily on some signals while ignoring others. More specifically, our results show that the average effect associated with the search action are marginal (mean effect size of 0.43\% on topic prevalence; not significant; $p > .05$). Interestingly, subscribing to a creator (the {\it Follow (U2I)} action) only has small effects on the homepage feed (mean effect size of 4.3\% on topic prevalence; significant). In contrast, Like and View actions have an average effect sizes of 17.5\% and 21.2\% on the topic prevalence.

\paraem{Liking videos results in drastic changes in homepage feed topic composition.}
Among all actions in our experiments, it appears that the \yt algorithm is very sensitive to users' likes. Examining our measurements of average treatment effect (\Cref{tab:topic-results}), we find his behavior is consistent and statistically significant across all topics and all our homepage topic composition vectors. This is also corroborated by the influence assigned to the {\it Like} variable by our constraint satisfaction problem (CSP) model, where it was found to be an order of magnitude more influential than any other action.
Specifically, the act of liking a video related to our treatment results in 13.8-34.4\% higher prevalence of that topic on the homepage. The changes are even more pronounced when considering the ordering of content on the homepage feed (mean effect size of 28.6\% on topic prominence), suggesting that the ranking algorithm does prioritize users' previous preferences. 

\paraem{Effects on homepage topic composition associated with subscribing to a creator are moderate.}
Interestingly, subscribing to a creator (the {\it Follow (U2I)} action), despite having a statistically significant effect, does not result in drastic homepage feed topic changes (mean effect size of 4.3\% on topic prevalence).
As we will see in \Cref{sec:rq2:results}, this is explained by the diversity of content produced by creators shown in \yt search results for `NFL' and `Fitness'.

\paraem{Homepage feeds are highly responsive to interactions with political content.}
Focusing once again on the like and view interactions, we see that the algorithm is very sensitive to political content. When our sockpuppets liked a political video, the prevalence and prominence of political content on their homepage feeds increased by 21\% and 34.4\%, respectively. This is once again corroborated by the value assigned to the `Politics' influence variable by our CSP model which attributed a 67\% and 637\% higher influence than was assigned to the `NFL' or `Fitness' variables, respectively.
There are several possible explanations for this behavior: (1) the prevalence of political videos shown to brand new users (the control group) is much lower than NFL- or fitness-adjacent content, making the post-treatment increase in political homepage feed content more dramatic; (2) the amount of fresh political content in the \yt library is high, leading the algorithm to recommend it generously to users falling in the `political' cluster after collaborative filtering; or (3) the creators of political content that were the subject of our sockpuppet interactions (the top creators returned for the search query `Elections') had less diverse content than creators of NFL and Fitness content (the top creators returned for the search query `Kansas City Chiefs' and `Fitness', respectively). As we will see in \Cref{sec:rq2:results}, some of these differences are explained by (3).

\para{\X.} In general, our results suggest that \X's homepage feed algorithms are rarely responsive to user actions and responses are topic-dependent. 

\paraem{User actions appear to have a moderate to low influence on the topics shown in the homepage feed.}
Only two actions, {\it Follow (U2I)} and {\it Like}, were found to have any statistically significant effect on the topics of the homepage feed, while no single topic resulted in a statistically significant change of the homepage feed composition. This finding is corroborated by our CSP model which shows a rather uniform influence across topics.
Further, even when the actions influence the topic composition of the homepage feeds, the effects were moderate. Specifically, the {\it Follow (U2I)} action only increased the prevalence and prominence of the corresponding topic by 2.6\% and 5.7\% (both are statistically significant), respectively, and {liking} content only increases its prominence (mean effect of 7.44\%; statistically significant). This suggests that our treatments with the {\it Like} interaction did not alter the content being presented, instead altering the ranking of the {liked} content.
We note that this is possible because of the selection of mainstream topics, already present in the default homepage feeds of platforms, in our experiments.
Given the mainstream nature of our selected topics, we can rule out the lack of related content or users as the explanation for our findings. There remain other possible explanations for this low responsiveness to user interactions. It is possible that the homepage algorithm is deliberately designed to not heavily personalize content for newer accounts with a small/non-existent network; or that the algorithm is slow to process user interactions for personalization, requiring more time for personalization than our experiments allowed; or that personalization on \X is generally not heavily driven by the user interactions we performed.
Unfortunately, the \X source code does not point to any of these explanations as correct because only a small subset of weights associated with engagement signals was released. Further, the released weights pertain to how the target content will be amplified rather than to the future personalization that a user will experience.

\para{Reddit.} Our results highlight several interesting features of Reddit's homepage feed algorithms.

\paraem{Community-centric recommendations.} Reddit's homepage feed algorithms appear to emphasize it's focus on community-centric engagement. This is evidenced by the high influence attributed to the {\it Join (U2C)} action. Joining a subreddit related to our treatment topics increased the prevalence and prominence of that topic by 21.7\% and 26.8\% (both statistically significant), respectively.  
This effect is much larger than that of the other actions used in our experiment and corroborated by our CSP model which assigned the U2C variable an influence value that was two orders of magnitude larger than any other action variable. This suggests that by subscribing to communities associated with their topics of interest, users have strong control over the content in their homepage feeds. 

\paraem{Opening posts and voting have a moderate influence on the topics of the homepage feed.}
In addition to the strong effects of U2C action, we also observed moderate effects associated with opening posts and upvoting them. The mean observed effect of opening posts related to a specific topic and upvoting them, on the topic prevalence vector, was 4.22\% and 3.26\%, respectively.
These effects were strongest in the NFL topic treatment groups (+10.75\% prevalence and +10.59\% prevalence, both statistically significant). Interestingly, these effects were much lower in the political and fitness treatments (<4\%). We hypothesize that this is an artifact of the specific clusters that our sockpuppets were placed in after collaborative filtering.

\paraem{Political topics have a drastic effect on the homepage feed only when you join their communities.}
It is worth noting in our results the oddity of the politics treatment effects. 
While other topics observed consistently high treatment effects (NFL) or consistently moderate treatment effects (Fitness), across the statistically significant actions (Open, Like, and U2C), our politics sockpuppets were inconsistent. 
They saw high treatment effects only after joining a political community and low-moderate effects otherwise. This suggests that simply viewing or voting on political content only has a marginal effect on the homepage feed, but joining a political community has a drastic effect on the homepage (+28.35\% topic prevalence; statistically significant). Given our understanding of Reddit's algorithms and the tendency to recommend more timely content, we hypothesize that this is an effect of the higher engagement in political communities than in NFL and Fitness communities. 

\para{Takeaway.}
The topics curated by homepage feed algorithms of \yt, \X, and Reddit reflect their distinct platform priorities and interpretations of user engagement signals. 
We see that \yt places a strong emphasis on the topics associated with direct user interactions such as liking and viewing videos. High amounts of topic-personalization for these actions highlight the platform's focus on immediate and explicit user feedback to personalize their experience and maximize watch times. Contrasting this with our finding regarding the moderate effects of subscribing to channels indicates that while subscriptions matter, real-time engagement is a more significant driver of the homepage feed algorithm. 
In comparison, \X appears to use a more measured approach where the topics shown users' homepage feeds appear to be less responsive to user interactions.
Meanwhile, Reddit emphasizes community-centric actions, with very high effects for joining subreddits. This focus on active community participation drives content promotion within specific interest groups, reflecting Reddit's structure.
Interestingly, we find that no platform appears to use users' {\it Search} queries for their homepage feed algorithms.
\section{How do user interactions influence platform behaviors?} \label{sec:rq2}

In \Cref{sec:rq1}, we described the influence that specific user interactions have on the topic composition of the homepage feed. Now, we focus on on learning about three specific behaviors of their homepage feed algorithms: (1) their exploitation and exploration tendencies; (2) their reliance on explicit association preferences; and (3) their dose-response behaviors. An overview of these behaviors is provided in \Cref{sec:background:homepage}. Here, we provide an description of our methods for measuring these behaviors (\Cref{sec:rq2:methods}) and then present the results of our analysis (\Cref{sec:rq2:results}).

\begin{table}[t]
    \centering
    \small
    \begin{tabular}{lp{4in}}
        \toprule
        \textbf{Notation} & \textbf{Definition} \\
        \midrule
        {\it ExploreProm}$(F, H^{\text{t}}, H^{\text{s}})$ & Refers to the prominence of content in feed $F$ that is unrelated to the sources or topics in the interaction history ($H^{\text{s}}$ and $H^{\text{t}}$). {\it [\Cref{eq:exploreprominence}]} \\
        {\it Topic(p)}  & Denotes the topic category associated with the post $p$. {\it [\Cref{eq:exploreprominence}]} \\
        {\it Source}$(p, H^{\text{sources}})$ & Denotes whether the post $p$ is categorized as `in-network' or `out-of-network' based on sources seen in prior interactions ($H^{\text{sources}}$). {\it [\Cref{eq:exploreprominence}]}\\
        {\it Rank(p, F)}    & Denotes the position rank of the post $p$ in feed $F$. {\it [\Cref{eq:exploreprominence}]}\\
        % \midrule
        %  $\mu_{t, a}^{\text{explore}}$ & The average prominence of exploratory content observed in feeds associated with (topic $t$, action $a$) interactions. {\it [\Cref{eq:avgintexplore}]}\\
        %  $\mu_{a}^{\text{explore}}$ & The average prominence of exploratory content observed in feeds associated with the action $a$. {\it [\Cref{eq:avgactexplore}]}\\
        %  $\mu_{t}^{\text{explore}}$ & The average prominence of exploratory content observed in feeds associated with the topic $t$. {\it [\Cref{eq:avgtopicexplore}]}\\
         \midrule
         {\it response(t, a, i)} & The treatment effect observed after the $i^{th}$ consecutive (topic $t$, action $a$) interaction on the platform. {\it [\Cref{eq:doseresponse}]}\\
        \bottomrule
    \end{tabular}
    \caption{Notation used in \Cref{sec:rq2:methods} along with their definitions and relevant equations.}
    \label{tab:rq2:notation}
\end{table}

\subsection{Methods for identifying platform behaviors} \label{sec:rq2:methods}
In this section, we describe how we use the homepage composition vectors (\cf \Cref{sec:methods:homepage}) to learn about platform behaviors. \Cref{tab:rq2:notation} describes the notation used in this section.

\para{Measuring exploration and exploitation tendencies.}
We characterize behaviors of the algorithm as exploratory when they recommend content unrelated to any prior user interactions with the platform. 
This is content that is neither related to the topics of prior interactions nor from the authors/creators that were the subject of prior interactions.
To measure exploratory behavior, we use the interaction histories of our sockpuppets to determine which topics and sources present in their homepage feeds {\em were not} the subject of any prior interactions with the platform. 
For example, for a sockpuppet which had undergone the fitness and politics treatment topics, we examined the prominence of NFL and Other content that were from sources characterized as `out-of-network' (\ie not the subject of any prior interactions).
Let $F$ be a homepage feed gathered from a sockpuppet with an interaction history including a set of topics ($H^{\text{topics}}$) and sources ($H^{\text{sources}}$). Then, we computed {\it ExploreProm} as:
\begin{align}
    {\it ExploreProm}(F, H^{\text{topics}}, H^{\text{sources}}) = & \sum_{\substack{(p \in F) \wedge (\textit{Topic}(p) \notin H^\text{topics}) \\ \wedge (\textit{Source}(p, H^{\text{sources}}) = \text{`out-of-network'})}} \frac{1}{{\it Rank(p) \times |F|}}
    \label{eq:exploreprominence}
\end{align}
Here, {\it Topic(p)} returns the topic category (NFL, Fitness, Politics, Cooking, or Other) associated with the post $p$ and {\it Source(p, $H^{\text{sources}}$)} returns the source type (in-network or out-of-network) based on the sources in the sockpuppet's interaction history ($H^{\text{sources}}$). 
This is a reasonable approximation of exploratory behaviors because they capture the prominence of homepage feed recommendations that are unrelated to any topics or sources contained in prior interactions. 
In our results, we report the distribution of the observed {\it ExploreProm} scores in response to each treatment. Specifically, for the (topic $t$, action $a$) interaction, we consider the {\it ExploreProm} scores across {\em all the homepage feeds} in our dataset whose interaction history included the treatment ($t$, $a$). 
This includes feeds where the full treatment sequence was not completed. For example, assume that a `Like' group sockpuppet was assigned the topic treatment sequence (NFL, Politics, Fitness) and we are measuring the exploration effects due to the interaction (NFL, Like). Then, the feeds recorded after each treatment (\ie NFL, NFL+Politics, NFL+Politics+Fitness) are included in our (NFL, Like) analysis. On the other hand, if we are measuring the exploration effects due to (Politics, Like), only the feeds recorded after the Politics treatment (\ie NFL+Politics, NFL+Politics+Fitness) are used for analysis.

\para{Measuring reliance on explicit association preferences.}
Explicit association signals such as Join (U2C) and Follow (U2I) differ from others because they do not require algorithms to interpret them. For example, it is unclear why a user liked a video --- was it due to the content within it, the creator that made it, or another unclear reason that the algorithm must interpret? In contrast, following or subscribing to a creator's channel is an explicit indicator from the user that they want more content from that creator.
In our analysis, we focus on highlighting the effects of the U2C and U2I interactions {\em on the sources of content} shown on users' homepage feed (\cf \Cref{tab:topic-results} and \Cref{sec:rq1:results} for effects of these actions on the topics of the homepage feed).
For this, we report the treatment effects of the U2C and U2I interactions on the {\it SourcePrevalence} and {\it SourceProminence} homepage composition vectors (\cf \Cref{eq:sourceprevalence,,eq:sourceprominence} in \Cref{sec:methods:homepage} for their definition). These are computed using the same approach as for the {\it TopicPrevalence} and {\it TopicProminence} vectors (shown in \Cref{sec:rq1:methods}). We then use the measured treatment effects on the homepage source vectors to characterize platforms' tendencies towards a network-dominant or algorithm-dominant information propagation model.

\para{Measuring dose-response effects.}
Recall that each interaction in a given treatment was repeated five times (on different search results each time), each 20 minutes apart (\cf \Cref{sec:methods:experiment} and \Cref{alg:crossover}). Our results thus far have reported the treatment effects associated with the first of these five interactions. We now use the remaining interactions to perform a dose-response analysis. 
For our analysis, we treated each repetition of the treatment (topic, action) interaction as a dose and measured the effects on the {\it TopicProminence} homepage composition vector caused by each subsequent (topic, action) treatment.
We used the measured treatment effects after each consecutive interactions (with the same topic and interaction, but different search result) as the platform's response. 
More concretely, for each (topic $t$, action $a$) pair, we compute:
\begin{align}
    {\it response(t, a, i)}  & = \hat{\mu}_{t, a}(i) \; ; \forall i
    \label{eq:doseresponse}
\end{align}
Here, {\it response(t, a, i)} denotes the platform's response to the $i^{th}$ successive dose of the ($t$, $a$) treatment and $\hat{\mu}_{t, a}(i)$ denotes the treatment effect of the $i^{th}$ application of the ($t$, $a$) treatment on the {\it TopicProminence} composition vector.
We then used the standard process for analyzing dose-response data by fitting the observed dose-response behavior to a sigmoidal function using the Hill equation \cite{Goutelle2008TheModelling}. We use characteristics of these curves to interpret platform behaviors across topics and actions.

\subsection{Results} \label{sec:rq2:results}

\begin{figure}[t]
    \centering
    \begin{subfigure}[b]{0.32\textwidth}
        \centering
        \includegraphics[width=\textwidth]{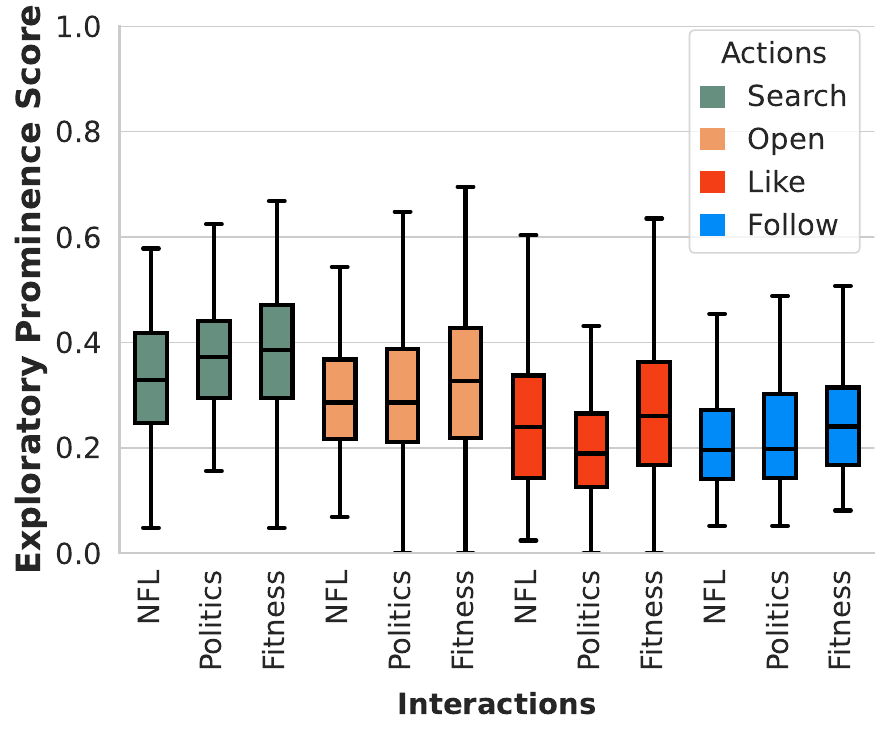}
        \caption{\yt}
        \label{fig:explore:interactions:yt}
    \end{subfigure}
    \hfill
    \begin{subfigure}[b]{0.32\textwidth}
        \centering
        \includegraphics[width=\textwidth]{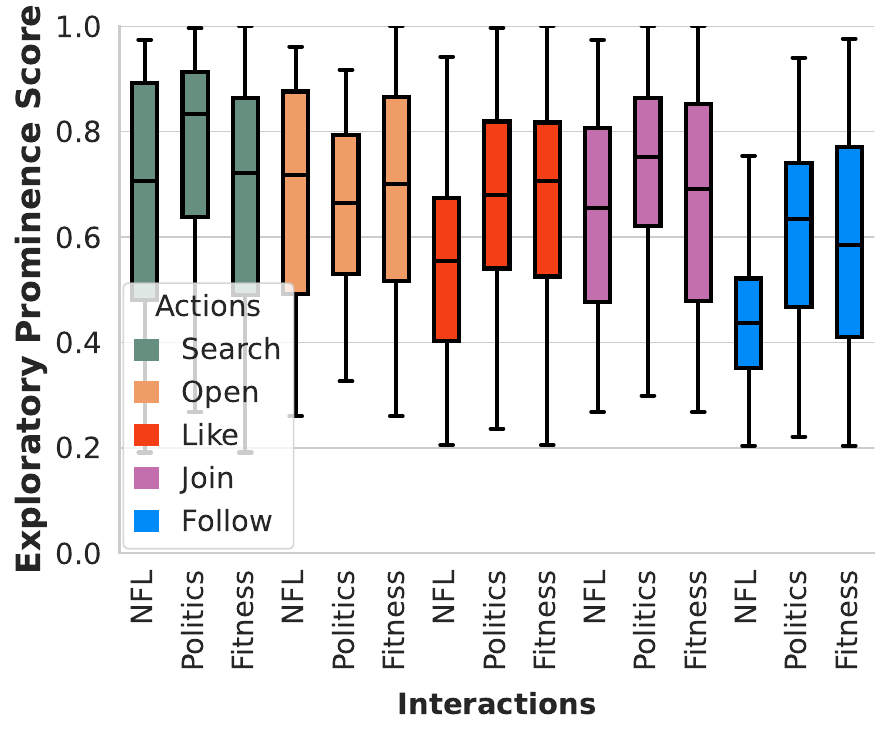}
        \caption{\X}
        \label{fig:explore:interactions:X}
    \end{subfigure}
    \hfill
    \begin{subfigure}[b]{0.32\textwidth}
        \centering
        \includegraphics[width=\textwidth]{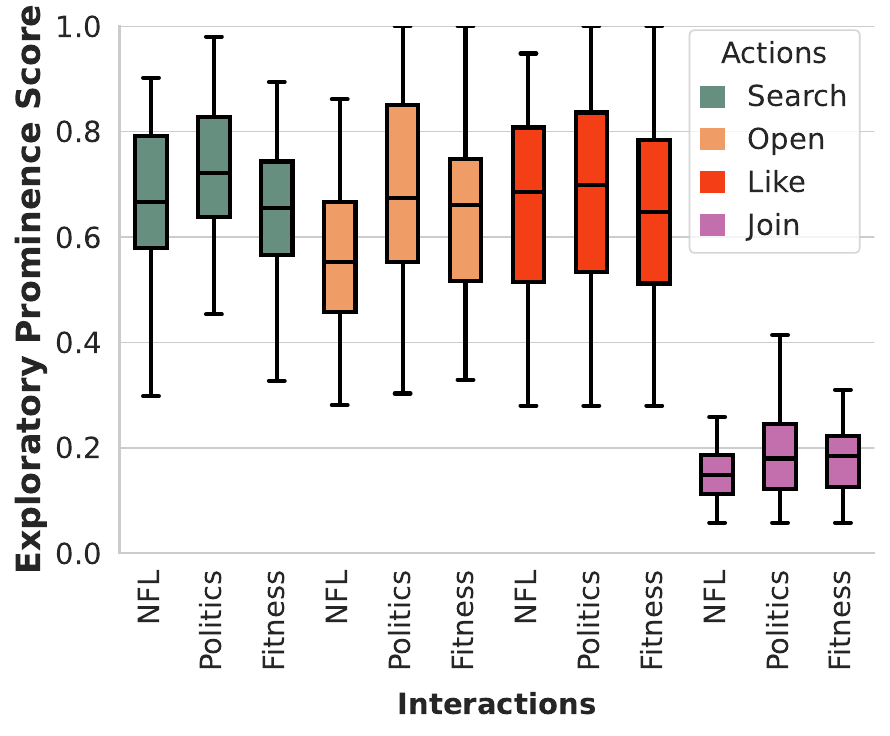}
        \caption{Reddit}
        \label{fig:explore:interactions:Reddit}
    \end{subfigure}

    \caption{Distribution of prominence of exploratory content on each platform in response to each treatment.}
    \label{fig:explore}
\end{figure}

\para{Exploration and exploitation tendencies.} A summary of our results are shown in \Cref{fig:explore}. They show the distribution of the prominence of exploratory content on users' homepage feeds on each platform, resulting from specific interactions on \yt (\Cref{fig:explore:interactions:yt}), \X (\Cref{fig:explore:interactions:X}) and Reddit (\Cref{fig:explore:interactions:Reddit}). From these we can make several interesting observations about the exploratory behaviors of each platform.

\paraem{Platforms' exploratory behaviors appear to be driven by prior user actions.}
We see that the variance in the prominence of exploratory content is high across actions and relatively low across each topic. 
This suggests that the homepage feed algorithms employed by \yt, \X, and Reddit are primarily action-driven and the topics associated with prior interactions are a secondary consideration.

\paraem{Consistency of measured effects on homepage feed exploratory behavior and topic composition.}
Consistent with our prior results in \Cref{sec:rq1:results}, we see that the homepage feeds of \X and Reddit are most reactive (\ie show less exploratory behavior) in response to the Follow (U2I) and Join (U2C) actions, respectively.
Contrary to our prior results on homepage topic composition which showed \yt's preference for the Like action over the Follow (U2I) action, we see that the algorithm does change its exploratory behavior more significantly in response to  Follow (U2I) action. 
Analyzing the data, we find that this is explained by the fact that the creators that our sockpuppets subscribed to were producers of more diverse content. This was most apparent for the NFL and Fitness topics.
Overall, we find that the \yt algorithm appears most responsive to the Open (viewing a video), Follow (subscribing to a creator), and Like actions, in that order. This seems consistent with \yt's public statements about the value of implicit signals (viewing videos and watch time) in their recommendation system \cite{Goodrow2021OnBlog}.

\paraem{\yt demonstrates more exploitative tendencies than other platforms.}
On average, across all topics and actions, \yt has the lowest prominence of exploratory content (mean {\it ExploreProm} score: 0.29). 
In contrast, the prominence of exploratory content was significantly higher on \X (mean {\it ExploreProm} score: 0.65) and Reddit (mean {\it ExploreProm} score: 0.53). In comparison to \yt, these differences were statistically significant. 
This suggests, in comparison with \X and Reddit, that the platform is more responsive to prior user interactions and content associated with the topics and creators of prior interactions are more prominent in users' homepage feeds.
Considering the exploration-exploitation tradeoff, \yt demonstrates significantly higher exploitative behavior than \X and Reddit.

\para{Reliance on explicit association preferences.}
Our results focusing on the exploratory behaviors of platforms' homepage curation algorithms highlight the importance of considering the sources of content (rather than only topics) on a homepage feed. 
This is underscored in our analysis regarding platforms' reliance on explicit association preferences summarized in \Cref{tab:source-results}. 
We make several interesting discoveries about platforms' homepage feeds and their reliance on explicit associations.

\begin{table}[h!]
    \centering
    \small
    \begin{adjustbox}{max width=\textwidth}
    \begin{tabular}{cclccccc}
        \toprule
        & & & \multicolumn{5}{c}{\textbf{Actions}} \\
        \cmidrule(lr){4-8}
        \textbf{Platform} & \textbf{Topic} & \textbf{Measure} & \textbf{$\hat{\mu}_{topic}$} & \textit{Open} & \textit{Like} & \textit{Join (U2C)} & \textit{Follow (U2I)} \\
        \midrule
        \multirow{12}{*}{YouTube} 
            & \multirow{3}{*}{NFL}      & {\it SourcePrevalence} & \textcolor{red}{1.11*} & 0.27 & \textcolor{red}{2.34*} & - & \textcolor{red}{1.82*} \\
            &                           & {\it SourceProminence} & \textcolor{red}{1.70*} & 0.73 & \textcolor{red}{4.55*} & - & \textcolor{red}{1.57*} \\
        \cmidrule(lr){2-8}
            & \multirow{3}{*}{Politics} & {\it SourcePrevalence} & \textcolor{red}{2.24*} & -5.24 & \textcolor{red}{7.41*} & - & \textcolor{red}{6.72*} \\
            &                           & {\it SourceProminence} & \textcolor{red}{6.95*} & -6.01 & \textcolor{red}{14.66*} & - & \textcolor{red}{19.12*} \\
        \cmidrule(lr){2-8}
            & \multirow{3}{*}{Fitness}  & {\it SourcePrevalence} & \textcolor{red}{1.63*} & -0.94 & \textcolor{red}{3.4*} & - & \textcolor{red}{4.13*} \\
            &                           & {\it SourceProminence} & \textcolor{red}{5.15*} & -1.3 & \textcolor{red}{9.83*} & - & \textcolor{red}{12.08*} \\
        \cmidrule(lr){2-8}
            & \multirow{3}{*}{\textbf{$\hat\mu_{action}$}}  & {\it SourcePrevalence} & - & -1.54 & \textcolor{red}{3.98*} & - & \textcolor{red}{3.91*} \\
            &                                               & {\it SourceProminence} & - & -1.72 & \textcolor{red}{9.01*} & - & \textcolor{red}{10.07*} \\
        \bottomrule
        \multirow{12}{*}{\X} 
            & \multirow{3}{*}{NFL}      & {\it SourcePrevalence} & \textcolor{red}{2.94*} & 0.0 & \textcolor{red}{5.34*} & 0.0 & \textcolor{red}{6.41*} \\
            &                           & {\it SourceProminence} & \textcolor{red}{7.81*} & 0.0 & \textcolor{red}{11.51*} & 0.0 & \textcolor{red}{19.71*} \\
        \cmidrule(lr){2-8}
            & \multirow{3}{*}{Politics} & {\it SourcePrevalence} & 0.24 & 0.0 & 0.22 & 0.0 & 0.73 \\
            &                           & {\it SourceProminence} & 0.68 & 0.0 & 0.06 & 0.0 & \textcolor{red}{2.67*} \\
        \cmidrule(lr){2-8}
            & \multirow{3}{*}{Fitness}  & {\it SourcePrevalence} & 0.69 & 0.78 & 0.23 & 0.0 & \textcolor{red}{1.73*} \\
            &                           & {\it SourceProminence} & \textcolor{red}{3.14*} & 0.48 & \textcolor{red}{3.63*} & 0.0 & \textcolor{red}{8.53*} \\
        \cmidrule(lr){2-8}
            & \multirow{3}{*}{\textbf{$\hat\mu_{action}$}}  & {\it SourcePrevalence} & - & 0.31 & \textcolor{red}{2.14*} & 0.0 & \textcolor{red}{2.78*} \\
            &                                               & {\it SourceProminence} & - & 0.19 & \textcolor{red}{5.70*} & 0.0 & \textcolor{red}{9.93*}\\
        \midrule
        \multirow{12}{*}{Reddit} 
            & \multirow{3}{*}{NFL}      & {\it SourcePrevalence} & \textcolor{red}{8.94*} & 0.0 & 0.0 & \textcolor{red}{26.83*} & - \\
            &                           & {\it SourceProminence} & \textcolor{red}{11.12*} & 0.0 & 0.0 & \textcolor{red}{33.4*} & - \\
        \cmidrule(lr){2-8}
            & \multirow{3}{*}{Politics} & {\it SourcePrevalence} & \textcolor{red}{12.43*} & 0.0 & 0.0 & \textcolor{red}{37.3*} & - \\
            &                           & {\it SourceProminence} & \textcolor{red}{13.49*} & 0.0 & 0.0 & \textcolor{red}{40.46*} & - \\
        \cmidrule(lr){2-8}
            & \multirow{3}{*}{Fitness}  & {\it SourcePrevalence} & \textcolor{red}{4.15*} & 0.0 & 0.0 & \textcolor{red}{12.5*} & - \\
            &                           & {\it SourceProminence} & \textcolor{red}{6.22*} & 0.0 & 0.0 & \textcolor{red}{18.68*} & - \\
        \cmidrule(lr){2-8}
            & \multirow{3}{*}{\textbf{$\hat\mu_{action}$}}  & {\it SourcePrevalence} & - & 0.0 & 0.0 & \textcolor{red}{24.02*} & - \\
            &                                               & {\it SourceProminence} & - & 0.0 & 0.0 & \textcolor{red}{30.31*} & - \\
        \bottomrule
    \end{tabular}
    \end{adjustbox}
    \caption{Influence of different (topic, action) pairs on the {\em `in-network' components of source composition vectors} of the homepage feed on \yt, \X, and Reddit. Each cell indicates the measured average $\hat{\mu}_{\text{topic, action}}$ treatment effect. $\hat{\mu}_{\text{topic}}$ and $\hat{\mu}_{\text{action}}$ denote the average treatment effect for each topic (across all actions) and action (across all topics), respectively. The {\it Search} interaction is excluded because it does not interact with content from any creators (all values are 0.0). Values marked with \textcolor{red}{red*} are statistically different from the control feeds ($p < .05$).}
    \label{tab:source-results}
\end{table}

\paraem{Platforms rely heavily on users' explicit association signals.}
Consistent with our prior analysis, we see that platforms generally do value explicit association signals. They appear to use these signals, more heavily than others such as Like and Open, to identify sources whose content should be promoted via homepage feeds. 
The effect is once again strongest on Reddit, a community-centric platform, where content from a joined community is between 18.6\% to 40.5\% more prominent on a user's homepage feed. This is in contrast to opening and upvoting content from a particular community, which both show no measured effect on the sources on a user's homepage feed.
On \X, we find that the Join (U2C) interaction is not used for homepage feed curation and the effects of U2I interactions are stronger than every other action (except for the curious case of Politics, which we discuss later).
On \yt, we find that the subscribe (U2I) interaction is weighed roughly equally to the Like interaction when determining sources of homepage content. However, viewing a creator's content by itself does not result in them becoming featured on the homepage feed. In fact, we see the opposite --- viewed creators become less featured on the homepage feed. 
This finding, paired with our results in \Cref{tab:topic-results} suggests that viewing a video is interpreted as being interested in the topic associated with the video, rather than content from that creator.

\paraem{Explicit association preferences heavily influence content ranking across all platforms.}
Generally, we find a large and significant difference between the effects on {\it SourcePrevalence} and {\it SourceProminence} for all topics and for all platforms (once again with the exception of political content on \X).
This difference is exacerbated for the explicit association signals in our experiments. 
This suggests that explicit associations are most heavily used to determine the ordering of content presented to users on their homepage feeds.
Taken together, this finding suggests that platforms are still more reliant on network-driven information diffusion than purely algorithm-driven diffusion. This is more clear in the case of \X and Reddit (where communities are part of the network).

\paraem{There are anomalous effects associated with sources of political content, relative to Fitness and NFL.}
Across all platforms, we notice that explicit associations with sources of political content have very different effects than the NFL and Fitness topic categories.
On \yt and Reddit, we find that these interactions result in the largest effects on homepage feed source prevalence and prominence composition vectors. This perhaps reflects the abundance of content from the associated creators and community.
On \X, however, we see that interactions with creators of political content on the platform did not result in their increased prevalence or prominence on users' homepage feeds for any interaction. This is despite the associated accounts having many other followers and tweets during the period of our study. 
This is a departure from the effects of interactions, implicit or explicit association, with NFL and Fitness content creators.
This deprioritization of political content (and their creators) on \X is also consistent with our findings in \Cref{tab:topic-results}.

\begin{figure}[t]
    \centering
    \begin{subfigure}[b]{0.32\textwidth}
        \centering
        \includegraphics[width=\textwidth]{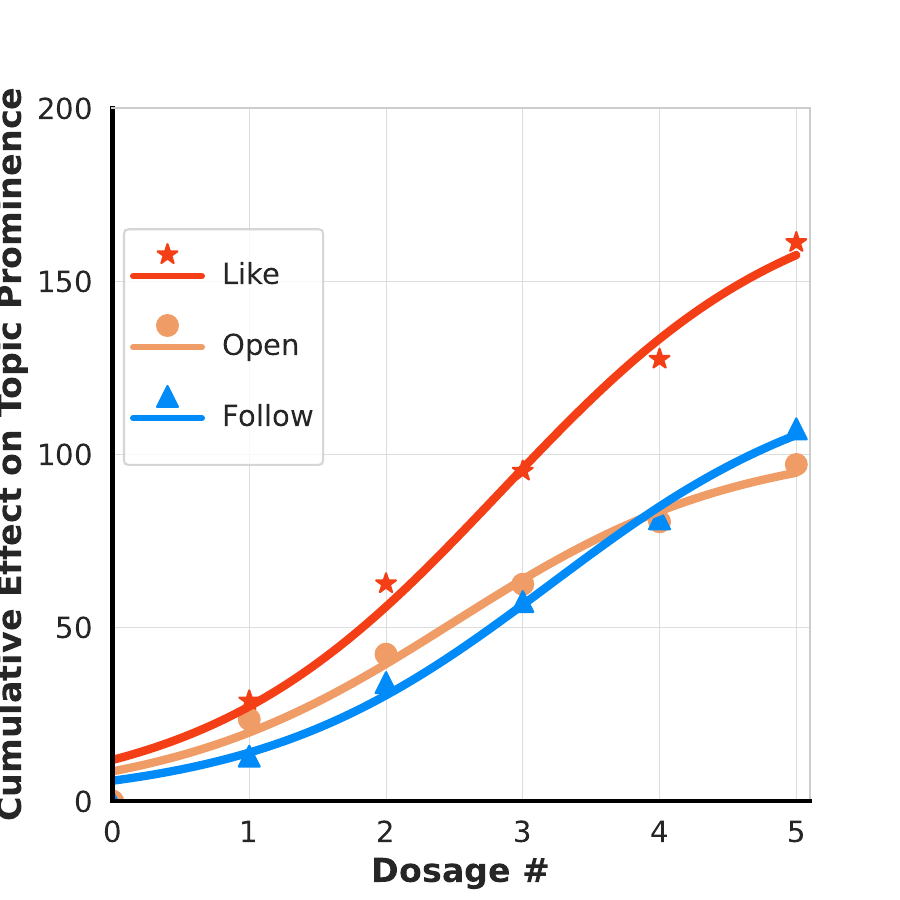}
        \caption{\yt}
        \label{fig:dose:action:yt}
    \end{subfigure}
    \hfill
    \begin{subfigure}[b]{0.32\textwidth}
        \centering
        \includegraphics[width=\textwidth]{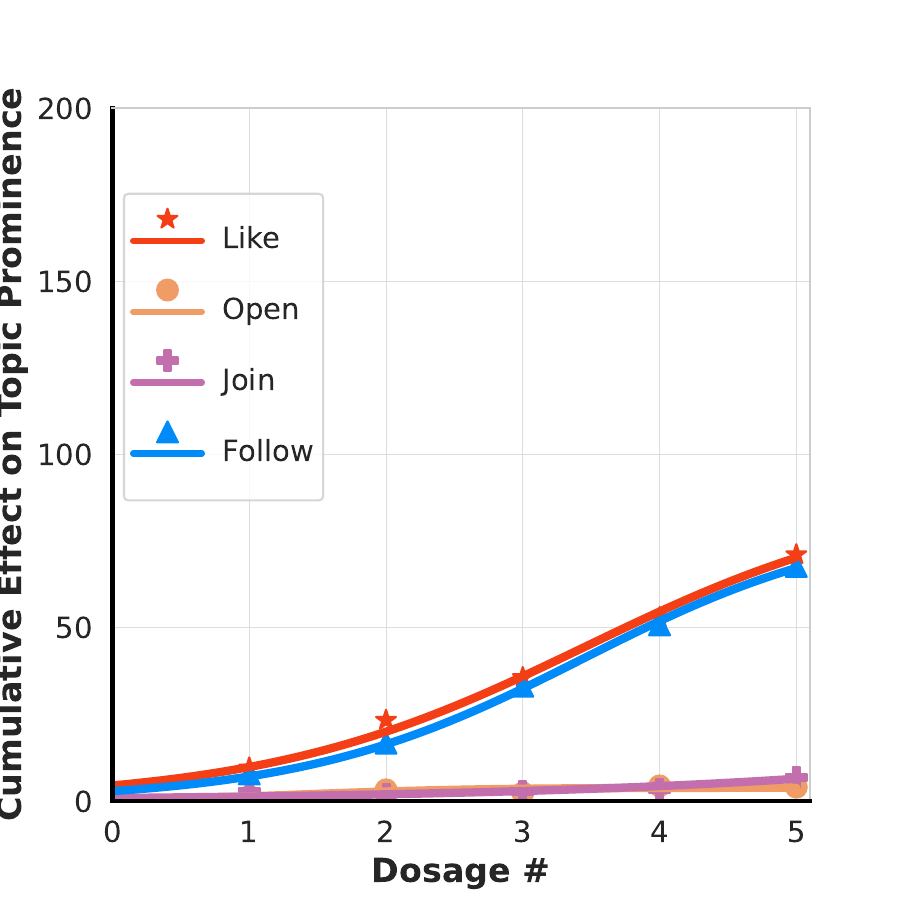}
        \caption{\X}
        \label{fig:dose:action:X}
    \end{subfigure}
    \hfill
    \begin{subfigure}[b]{0.32\textwidth}
        \centering
        \includegraphics[width=\textwidth]{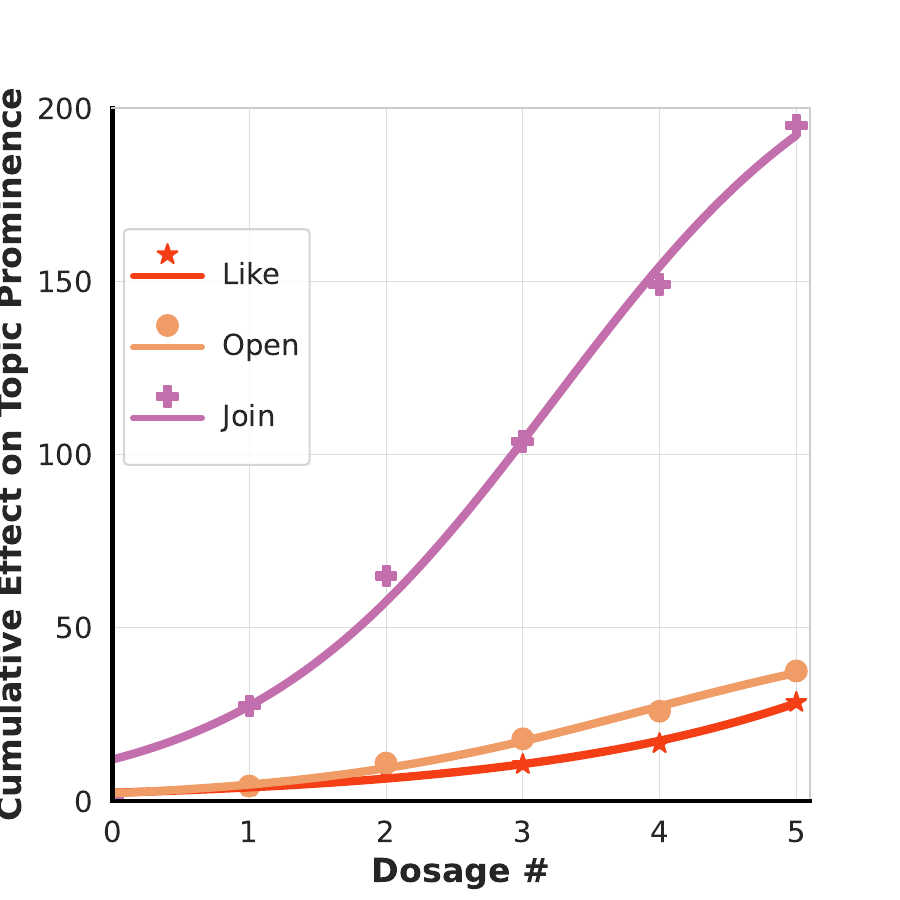}
        \caption{Reddit}
        \label{fig:dose:action:Reddit}
    \end{subfigure}

    \begin{subfigure}[b]{0.32\textwidth}
        \centering
        \includegraphics[width=\textwidth]{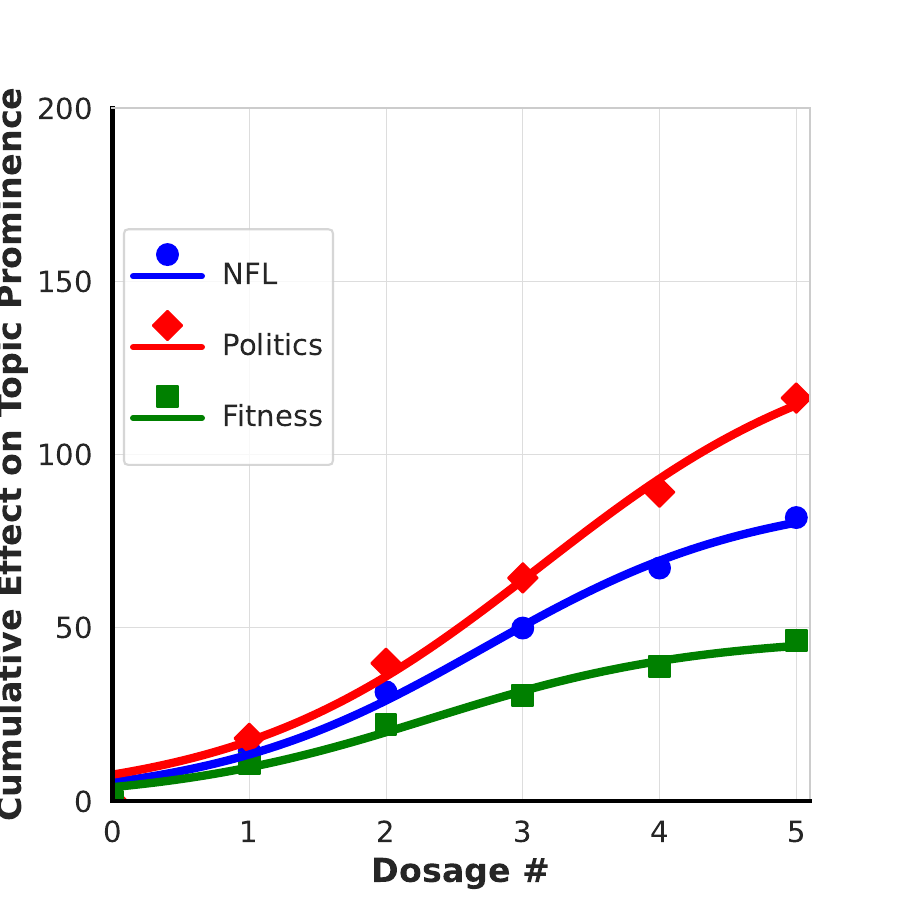}
        \caption{\yt}
        \label{fig:dose:topic:yt}
    \end{subfigure}
    \hfill
    \begin{subfigure}[b]{0.32\textwidth}
        \centering
        \includegraphics[width=\textwidth]{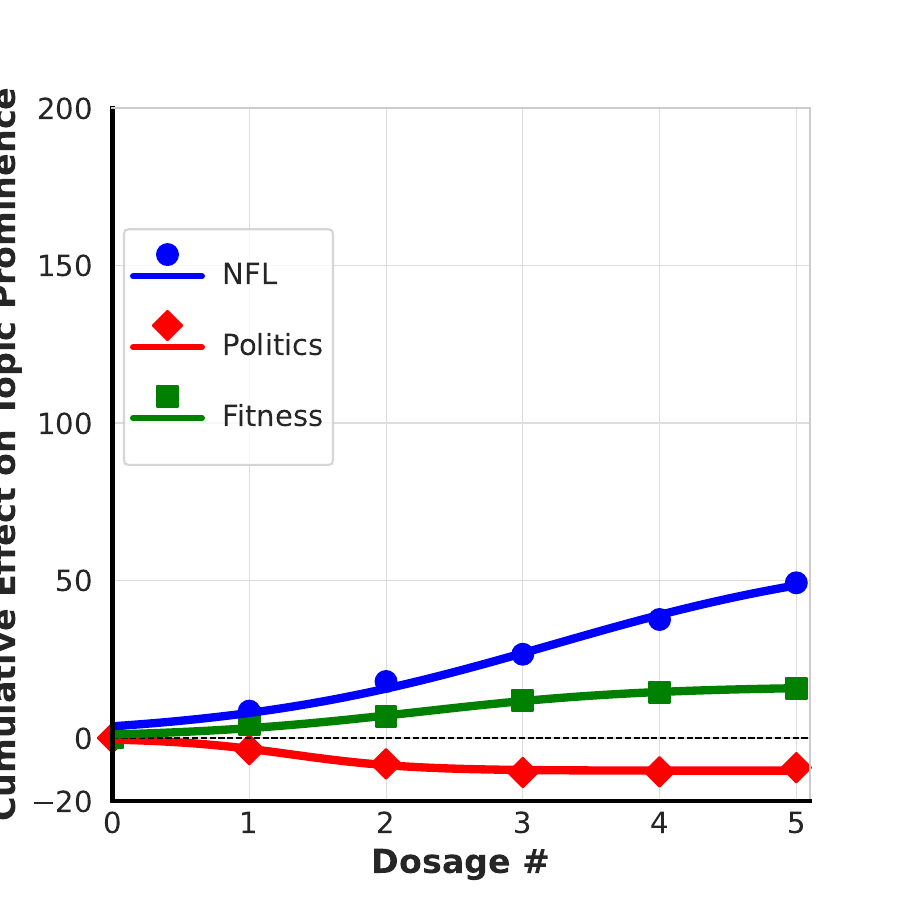}
        \caption{\X}
        \label{fig:dose:topic:X}
    \end{subfigure}
    \hfill
    \begin{subfigure}[b]{0.32\textwidth}
        \centering
        \includegraphics[width=\textwidth]{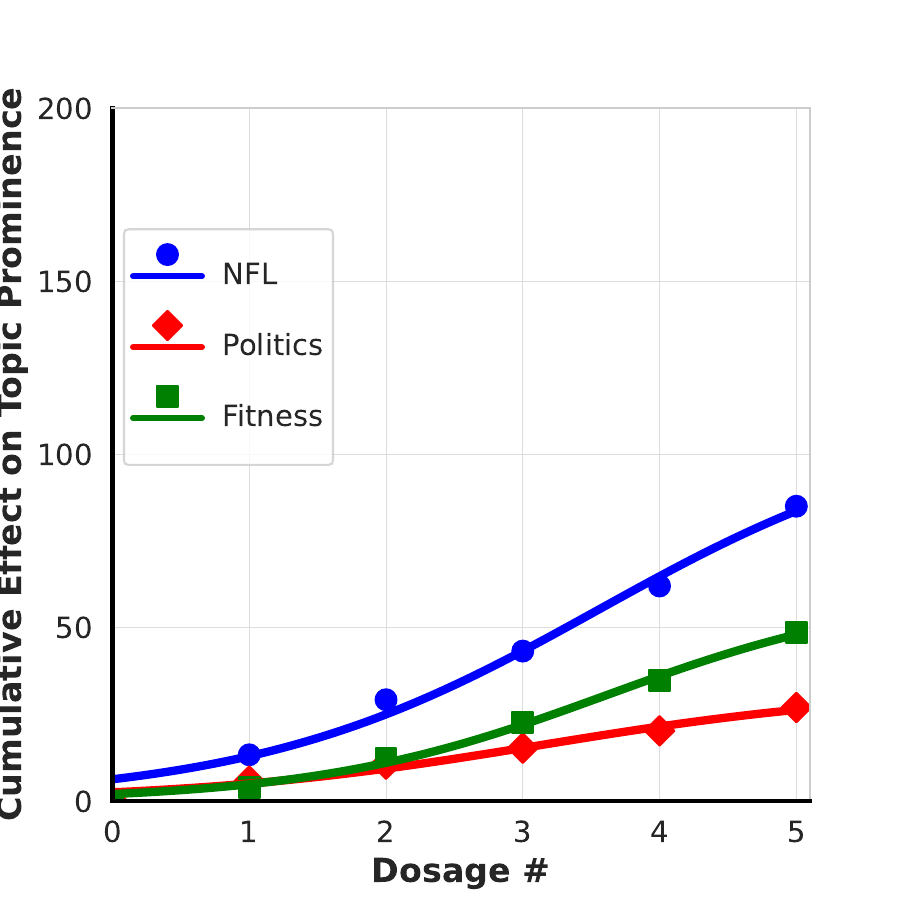}
        \caption{Reddit}
        \label{fig:dose:topic:Reddit}
    \end{subfigure}

    \caption{Cumulative effects of consecutive nearly-identical (topic, action) interactions on the {\it TopicProminence} homepage composition vector of each platform. \Cref{fig:dose:action:yt,,fig:dose:action:X,fig:dose:action:Reddit} show the average effects from each action, across all topics and \Cref{fig:dose:topic:yt,,fig:dose:topic:X,,fig:dose:topic:Reddit} show the average effects from each topic, across all actions. A Sigmoid fit along with the mean square error is shown for each curve.} 
    \label{fig:dose}
\end{figure}

\para{Dose-response behaviors.} A summary of our findings about the effects of multiple consecutive (topic, action) interactions (\ie multiple doses) on {\it TopicProminence} homepage composition vectors (\ie the response), broken down by actions and topics for each platform, is illustrated in \Cref{fig:dose}. \Cref{fig:dose:action:yt,,fig:dose:action:X,fig:dose:action:Reddit} show the average effects of each action and \Cref{fig:dose:topic:yt,,fig:dose:topic:X,,fig:dose:topic:Reddit} show the average effects of each topic. We identify several interesting trends related to platforms' dose-response behaviors.

\paraem{\yt prioritizes the Open action less over time.}
Examining \yt's response behaviors broken down by specific actions (\Cref{fig:dose:action:yt}), we see that the Like, Follow (U2I), and Open actions all continue to play an important role in the prominence of topics for users' homepage feeds, even when they are applied in succession.
They appear to influence the homepage in a near linear fashion --- \ie consecutive nearly identical interactions, with the same (topic, action), each result in approximately the same effects on the prominence of topics on the homepage feed.
Examining the slopes of these increases, we find that Liking videos consistently shows the strongest effects on the prominence of topics. 
Interestingly, we see that despite having a higher initial influence on the homepage feed than the Follow (U2I) action, the slope associated with the Open action is lower --- \ie it's influence on the prominence of topics in the homepage feed becomes lower than the Follow action after multiple interactions.
This could be evidence that the \yt algorithm becomes more more confident in identifying the subject of a users interest when they use the Like interaction, rather than the Open interaction.
Examining the response behaviors broken down by specific topics (\Cref{fig:dose:topic:yt}), we find that repeated interactions with political content continues to yield a consistent and high effect on the homepage. 
This is in contrast with NFL and fitness content which have a declining influence after multiple interactions. 
We hypothesize that this effect is due to the large amounts of high-engagement and timely politics-related content available on \yt. 

\paraem{\X continues to show anomalous behavior for interactions with political content.}
Consistent with our earlier results about the deprioritization of political content, our analysis of \X's dose-response behaviors broken down by topic (\Cref{fig:dose:topic:X}) once again showed that interactions with political content were treated differently than interactions with NFL- or fitness-related content. 
Rather surprisingly, we found that the influence actually turned negative after multiple interactions with political content (\ie the control group, which had no political interactions, had more political content than the treatment group). 
Based on this analysis and our prior results, we hypothesize that \X has unique algorithmic responses which reduce a users exposure to political content, even after the explicitly seek such content. It remains unclear whether this is intentional.
In our analysis of the influence of actions on \X's homepage feed \Cref{fig:dose:action:X}, we find that the influence of Like and Follow actions have a similar influence on the prominence of topics on the homepage, even after multiple consecutive interactions.

\paraem{On Reddit, up-voting and viewing interactions gain importance over time.}
Examining Reddit's dose-response behavior aggregated by actions (\Cref{fig:dose:action:Reddit}), we see that the effects of the Join (U2C) interaction on topic prominence remain consistent even when they are repeated in succession. Once again, this demonstrates Reddit's structure as a collection of interest-based communities and the algorithm's understanding of a Join action as a strong and explicit indicator of a users interest.
Interestingly, we see that the prominence of topic related to posts that a user views or up-votes become higher after multiple consecutive interactions. We hypothesize that Reddit's homepage feed algorithm gains increasing confidence in users' topic preferences indicated through less explicit signals over time --- perhaps due to the more effective collaborative filtering that occurs after multiple interactions.

\para{Takeaways.}
Our analysis of exploratory tendencies shows that platforms differ in how they navigate the explore-exploit trade-off. We find that \yt is significantly faster at personalizing the content of their homepage feeds in response to prior user interactions than \X or Reddit. We see that this behavior is primarily driven by prior actions rather than topics, which have only a secondary influence on the prominence of exploratory homepage feed content.
We discover that all platforms rely heavily on explicit association preferences from users when curating the homepage feed and that content associated with these signals are prioritized by all platforms' ranking algorithms.
In our dose-response analysis, we find that algorithms appear to gain more confidence in their understanding of user preferences after multiple similar interactions. In some cases, this eventually causes a reordering of the influence of specific actions on the homepage feed.
Finally, across multiple analyses, we discover evidence of deprioritization of political content on \X. 

\section{Discussion} \label{sec:discussion}
We now discuss the limitations of our study (\Cref{sec:discussion:limitations}), highlight the need for increased transparency from platforms (\Cref{sec:discussion:transparency}), and describe a promising direction for future research (\Cref{sec:discussion:implications}).

\subsection{Limitations} \label{sec:discussion:limitations}
This research is a `best-effort' measurement study aimed at bringing transparency to the homepage feed curation algorithms employed by \yt, \X, and Reddit. Like all other research, it comes with limitations. We describe these below and highlight how they may have influenced our findings.

\para{Confounders due to crossover trial design.}
Our study relied on a crossover trial design due to the significant scalability advantages it provides. However, it also introduces potential confounders that could affect the accuracy and reliability of our findings. 
One major issue with a crossover design is carryover effects, where interactions from prior treatments can influence the measurements of subsequent ones, thereby introducing noise into our data. 
Prior to data collection, we rigorously tested several strategies to mitigate these effects and ultimately implemented the most effective method, involving washout periods and interactions between each treatment and comparisons with paired controls.
Despite these efforts, we expect that some residual confounding may still persist. Nevertheless, we believe, based on manual observations and statistical testing (\cf \Cref{sec:rq1:methods}), that any such confounding is marginal and unlikely to impact the overall conclusions of our study.

\para{Limited interactions and topics.}
Our study concentrated on measuring the effects associated with five types of non-intrusive user interactions across three carefully selected topics for each platform. 
This focus was necessitated by the high costs associated with creating new accounts for each interaction (and platform) and the technical challenges of synchronizing interactions across all sockpuppets to minimize temporal effects.
While we believe that our approach is methodologically sound for uncovering platform responses to our specifically chosen topics and actions, we recognize that the identified behaviors may not fully represent the platforms.
For example, our selection of mainstream topics means we might miss platform behaviors that become apparent only with less common and niche content. Similarly, our choice of interactions, driven by practical and ethical considerations, excluded frequently utilized features such as negative feedback, sharing, and content creation affordances.
Furthermore, by limiting our dose-response analysis to five interaction doses, we may not have fully captured additional effects and behavioral changes that emerge as account activity and age increase. 
Despite these limitations, we are confident that our findings provide valuable insights into platform behaviors within the scope of our study.

\para{Focus on text data.}
Our study focused exclusively on the textual elements of the homepage feed, ignoring video and image elements. This decision was driven by the high costs, impreciseness, and validation challenges associated with models for video and image analysis. Consequently, we may have missed nuances conveyed through video or image posts on homepage feeds. 
Nonetheless, we believe this is unlikely since text plays a crucial role in contextualizing video and image elements through titles, descriptions, and captions. By leveraging Large Language Models (LLMs) to generate text embeddings, we obtained embeddings that capture a deeper contextual understanding of the posts compared to conventional token-based analysis. 

\para{Choice of platforms.} We acknowledge that our study omits several popular platforms, including TikTok, Facebook, and Instagram. Although the popularity of these platforms with specific demographics warrants their inclusion in a study like ours, practical considerations made their evaluation infeasible. 
Specifically, we encountered significant challenges in engineering sockpuppets that could consistently interact with these platforms due to difficulties in creating new accounts and the frequent blocking of accounts shortly after their creation. This limited our ability to include these platforms in our analysis, but we believe the platforms we did study still provide valuable insights into social media behaviors.

\subsection{Yet another call for platform transparency}\label{sec:discussion:transparency}
By mediating and influencing our communications and information consumption, social media platforms wield enormous power and influence over society. Despite this influence, their operation remains opaque to the public. It is therefore unsurprising that there have been many calls for increased transparency from platforms. 
These calls have emerged from a variety of stakeholders including users, regulators, and researchers and focus on a range of issues including platform governance and moderation, data privacy practices, and content curation algorithms \cite{Jiang2019BiasModeration, McCarthy2021HowBrookings, Krishnan2023HowInstitute, Jozwiak2022WhatInitiative, Rep.Trahan2022H.R.67962022, Sen.Coons2023S.1876Act, Davidson2023Platform-controlledScience}. 
In the context of transparency of content curation algorithms, despite a large amount of spent resources, platform users and researchers still lack a comprehensive understanding of exactly how platforms curate content and how user behaviors feed into this process. 
In a positive step forward, some platforms have recently made minor concessions by providing explanations for why users see certain content \cite{2022LearnNewsroom}. Unfortunately, these explanations are often too general to be useful to their users or researchers \cite{Mousavi2024AuditingExplanations}. 
While this study and many others have shed light on previously opaque content curation processes, they are far from providing a complete understanding due to  `artificial' (\ie non-scientific) reasons that arise solely from platform-imposed restrictions. 
For instance, in this study, we faced substantial challenges due to denial of access to platform resources (\eg accounts that could be used for an external audit without being blocked). As a result, we had to expend significant resources simply to create accounts for each platform and engineer sock puppets that could circumvent the platforms' bot detection techniques. This restriction forced us to examine only a limited number of interactions and to utilize a sub-optimal experimental design to achieve scalability.
The current situation where platforms select and collaborate with their own auditors underscores the power asymmetry that exists between platforms and stakeholders. 
We urge lawmakers to continue pursuing legislation to facilitate independent and unbiased audits and assessments of platform behaviors \cite{Sen.Coons2023S.1876Act, Rep.Trahan2022H.R.67962022}. Such legislation is crucial for enhancing platform accountability and mitigating societal harms.

\subsection{Directions for future research: A focus on the study of platform behaviors}\label{sec:discussion:implications}
While platform algorithms are built on seemingly simple and predictable objectives, their interactions with other algorithms, users, and platforms lead to complex and opaque emergent tendencies that often defy straightforward analysis. Borrowing from the language of Rahwan et al. \cite{Rahwan2019MachineBehaviour}, we refer to these tendencies as {\em platform behaviors}. Identifying these behaviors requires characterizing platforms’ responses to stimuli in a broader context and from more diverse perspectives than current audits typically employ.
For instance, by examining the explore-exploit tendencies of a platform, we can better understand and explain why it may heavily recommend problematic content (\eg extremist, conspiratorial, age-inappropriate) to users who show an interest in such topics. These characterizations are not only useful but necessary to fundamentally understand: (1) how platforms operate, (2) platforms' influence on their users, and (3) how societal harms caused by platforms may be mitigated.
Ignoring these higher-level characterizations of platform behaviors likely means that audits will continue to illuminate only singular aspects of platforms, thereby limiting their usefulness and relevance. Therefore, we propose that researchers work towards the development of a taxonomy of platform behaviors (and mechanisms for measuring them). This idea draws inspiration from the field of psychology, where the APA’s Diagnostic and Statistical Manual (DSM) \cite{AmericanPsychiatricAssociation2022DSMDisorders} was transformational in introducing a shared language and, with it, a broader understanding of human behaviors, disorders, and their diagnosis.
We acknowledge the challenges posed by limited access to platforms and the high costs of audits. Nevertheless, given the potential high payoffs, it is crucial that researchers explore this avenue.

\section{Related Work} \label{sec:related}
% Social media platforms and search engines employ sophisticated algorithms to curate and present information to users. These algorithms primarily leverage user interactions and content understanding to personalize experiences. As algorithmic curation has become increasingly central to information consumption, researchers have developed various methods to study its impact and potential consequences. Due to the opaque nature of these algorithms and limited access to internal data, researchers primarily employ observational or experimental study designs. Observational studies analyze real user data but are often limited by sample size and representativeness. Experimental studies, using techniques like sock puppet methodology, emulate user behavior to observe algorithmic responses. However, platforms actively deter such studies, presenting additional challenges. Despite these limitations, researchers have made significant progress in understanding platform behaviors and their potential societal impacts.

In this section, we highlight the influence of prior platform audit efforts on our study and put our work in the context of their results.

\para{Audits to understand platform interpretations of user interactions.}
Across various platforms, researchers have investigated how different user interactions shape content recommendations. These studies provide insights into the mechanics of recommendation algorithms and their responsiveness to user interactions.
Focusing on \yt, research by Liu \etal \cite{liu_how_2024} found that watching videos had a significant positive effect on homepage recommendations. Our research replicates this finding. They also noted an increased effectiveness of the ``{\em not interested}'' button in removing unwanted recommendations. This was in contrast to earlier findings which showed that these mechanisms were ineffective for controlling video recommendations \cite{Kiros2022YouTubesReview}. 
Chandio \etal \cite{Chandio2024HowSystems}, while examining the reasons for contradictory findings in prior \yt audit research, identified that video watch patterns and orders influenced the content recommended to users via the sidebar recommendations. Our analysis regarding the influence of the Open action for the homepage feed compliments this finding.
Our study was also inspired by the work of Vombatkere \etal \cite{Vombatkere2024TikTokFeeds} who developed a framework to analyze TikTok's home feed composition and personalization using a combination of sock puppet studies and real user data. They found that liking and following interactions were the most influential factors in shaping the home feed, confirming earlier findings by Boeker \etal \cite{Boeker2022AnTikTok}. They also examined the explore-exploit trade-off on TikTok and found that 30-50\% of content on the home feed contained potentially exploitative recommendations, raising concerns about the platform's content curation practices. Our study finds that these exploitation behaviors are comparable to \X and Reddit, but far less prominent than \yt.

\para{Audits to examine downstream societal effects of platform use.}
Over the past few years, there has been a significant effort to understand how platforms and their recommendation systems may contribute to harms in society --- particularly focusing on the spread of extremist content and polarized ideologies.
On \yt, early sock puppet studies found the platform disproportionately recommending fringe far-right content \cite{Ribeiro2020AuditingYouTube}. This finding provided the foundation for later audits which confirmed the presence of filter-bubble effects on \yt \cite{Haroon2022YouTubeRecommendations, SrbaIvan2023AuditingBubbles}. 
However, more recent studies using a combination of synthetic and real user traces have found that the consumption of radical and extremist content is driven more by user behavior than algorithmic amplification \cite{hosseinmardi_examining_2021, Chen2023SubscriptionsChannels}. 
This shift in findings can be attributed partly to methodological advancements and, crucially, to changes in \yt's recommendation algorithm. 
On \X, through a sock puppet study emulating real-world users, researchers found that the platform's algorithm increased source diversity while exacerbating partisan differences. Additionally, Huszár \etal discovered algorithmic amplification favoring right-leaning news sources \cite{huszar_algorithmic_2022}. This was supported by prior research from Chen \etal \cite{chen_neutral_2021} who observed higher polarization and potential filter bubbles associated with conservative accounts. 
Reddit, with its unique structure of topic-based communities, has been the birthplace of various problematic communities and ideologies \cite{massanari2017gamergate, habib_act_2019, habib_making_2022}. However, because of the network-driven nature of Reddit, much research has focused on examining the downstream effects of their platform governance strategies \cite{chandrasekharan2019crossmod, Chandrasekharan-CSCW2017, habib_exploring_2022}.
Notably, research has shown that Reddit's network-driven nature results in fewer echo chambers, compared to other platforms \cite{DeFrancisciMorales2021NoReddit, Cinelli2021TheMedia} and prior findings of high polarization can be explained by the influx of polarized users rather than filter-bubbles \cite{Waller2021QuantifyingPlatforms}.
Recent studies conducted in collaboration with Meta indicated that the News Feed algorithm may actually lower exposure to misinformation compared to a chronologically sorted feed and highlighted the role of the homepage feed algorithm for shaping diverse content exposure \cite{Guess2023HowCampaign, Guess2023ResharesOpinions}. 
Although our study {\em does not} study the downstream effects of recommendations, our findings regarding platforms' explore-exploit, explicit association, and dose-response behaviors explain many of these previously discovered artifacts.

\section{Concluding Remarks} \label{sec:conclusions}
This study provides several valuable insights about the influence of user interactions on homepage feeds and platform behaviors.

\para{Homepage feed algorithms reflect platform priorities.}
The homepage feed functions as the central interface for user engagement and information consumption. How the information is curated within the homepage reveals the platform's objectives and structures. Our findings highlight that YouTube, \X, and Reddit each interpret user interactions differently. 
YouTube demonstrates high responsiveness to explicit engagement signals such as likes and views, aligning with its focus on maximizing watch time and user engagement. In contrast, \X exhibits a more measured approach, with homepage feeds showing lower responsiveness to user interactions overall. Reddit emphasizes its community-centric nature, with the Join interaction having the most significant impact on homepage content curation.

\para{Topics take a secondary role to actions in the curation of homepage feeds.} Across all studied platforms, we find that user actions play a more substantial role than topics in determining homepage feed content. The specific interactions a user performs (e.g., liking, viewing, following) have a greater impact on future recommendations compared to the topical content associated with those interactions. This suggests that platforms prioritize understanding how users engage rather than solely focusing on what content they engage with.

\para{Implicit signals gain influence over time.} Our analysis reveals that as users interact more with a platform, algorithms appear to become more adept at interpreting less explicit signals. For instance, on Reddit, we observed that the impact of viewing or upvoting content increases after multiple interactions, indicating that the algorithm refines its ability to infer user preferences from these subtle cues over time.

\balance

%\footnotesize

\bibliographystyle{ACM-Reference-Format}
\bibliography{references}

%\newpage

%\input{appendix}

\end{document}